\documentclass[12pt]{article}
\usepackage[colorlinks=true,citecolor=blue,linkcolor=blue]{hyperref}
\usepackage{multirow}
\usepackage[left=2cm,right=2cm,top=2cm,bottom=2cm]{geometry}
\usepackage{cite}
\usepackage{psfrag}
\usepackage[demo]{graphicx}
\usepackage{graphicx}
\usepackage{graphics}
\usepackage{epsfig}
\usepackage{booktabs}
\usepackage{amsmath,amsfonts,amssymb}
\usepackage{pstcol,pst-fill,pst-grad}
\usepackage{pstricks,pst-fill,pst-grad}
\usepackage{euscript}
\usepackage{pstricks}
\usepackage{wrapfig}
\usepackage{slashed}
\usepackage[toc,page]{appendix}
\usepackage{here}

\begin{document}
	\title{\vspace{-3cm}
		\hfill\parbox{4cm}{\normalsize \emph{}}\\
		\vspace{1cm}
		{ Laser-assisted neutral Higgs-boson pair production in Inert Higgs Doublet Model (IHDM)}}
	\vspace{2cm}
	
	\author{ M. Ouhammou,$^1$ M. Ouali,$^1$  S. Taj,$^1$  Y. Attaourti,$^2$ and B. Manaut$^{1,}$\thanks{Corresponding author, E-mail: b.manaut@usms.ma} \\
		{\it {\small$^1$ Sultan Moulay Slimane University, Polydisciplinary Faculty,}}\\
		{\it {\small Research Team in Theoretical Physics and Materials (RTTPM), Beni Mellal, 23000, Morocco.}}\\
		{\it {\small$^2$ High Energy Physics and Astrophysics Laboratory, FSSM, UCAM, Morocco.}}\\				
	}
	\maketitle \setcounter{page}{1}
\date{}
\begin{abstract}
In the framework of the Inert Higgs Doublet model (IHDM), we have investigated, in the center of mass frame, the neutral Higgs-boson pair production in the presence of an intense and circularly polarized laser field via ${e}^{+} {e}^{-}$  annihilation $({e}^{+} {e}^{-}\rightarrow H^{0}A^{0})$ at the lowest order. By using the scattering matrix method, we have derived the analytical expression of the differential cross section. The latter is numerically integrated over the solid angle to obtain the total cross section. Then, we have shown how this total cross section depends on the outgoing particles mass for different number of exchanged photons. Next, we have illustrated its variation as a function of the centre of mass energy for different neutral Higgs-boson masses. Finally, we have indicated how it changes as a function of the laser field amplitudes for both different number of exchanged photons and different neutral Higgs-bosons masses.
\end{abstract}
\maketitle
\section{Introduction}
After the intervention of the laser almost five decades ago, enormous technological advances have supplied experimentalists with extremely short and powerful pulses of coherent electromagnetic radiation \cite{1,2}. Physicists have been motivated to investigate in a growing variety of different branches of physical research. Recently, much attention was devoted to the possibility of using powerful laser fields to test physical theories such as standard model and beyond \cite{3}. This interest is due not only to the rapid development of laser sources but also for the fact that a large of unknown phenomena were induced by laser-matter interaction \cite{4,5,6,7}. In general, particle physics' interaction can be divided into two types. The first type exist also in the presence of the laser field, yet it can be modified in its presence \cite{8}. The second type is laser-induced processes which can not occur in the absence of an external field \cite{9}. Several weak decay and scattering processes were studied. In \cite{10,6}, it is found that the laser field has a great impact on the decay processes as it modify the particles lifetime and affect its decay modes. In our recent work \cite{11} we have found that the total cross section of the Higgs-strahlung process can be largely reduced by several orders of magnitude only in extreme intense electromagnetic fields.

Higgs-boson, Which is discovered in 2012 by ATLAS and CMS experiment with a mass of approximately 125 GeV \cite{12,13}, plays an important role in symmetry breaking mechanism in the standard model. Based on the data of Run-1 and Run-2, several Higgs-boson properties, which are in full agreement with the prediction of the standard model, have been measured at the Large Hadron Collider (LHC) \cite{14,15,16,17,18,19}. Therefore, the standard model is recognized as a very successful model even though there are strong hints of new physics beyond the standard model. Then the latter could not be an ultimate theory but, it should instead be viewed as a low energy effective theory of some more complete and fundamental one yet to be established.
After the discovery of the Higgs-boson, there have been several theoretical and phenomenological studies which are devoted to non-minimal Higgs sector models that can explain such discovery and address some of the standard model's weakness. One of the simplest non-minimal Higgs model is the popular IHDM which is a version of 2HDM with an exact discrete $Z_{2}$ symmetry \cite{20}. It provides a viable dark matter candidate as the standard model scalar sector parametrized by $H_{1}$ and extended by an inert scalar doublet $H_{2}$.

It is well known that one of the main goals of the future run of the Large Hadron Collider (LHC) is to improve its measurement. Moreover, it is expected that such a precise measurement can be performed at the future electron positron ($e^{+}e^{-}$) collider such as the Circular Electron Positron Collider (CEPC) \cite{21}, the Compact Linear Collider (CLIC)\cite{22}, the Future Circular Collider (FCC-ee) \cite{23} and the International Linear Collider (ILC)\cite{24}. These ($e^{+}e^{-}$) colliders, which possess a very clean environment, were expected to deliver high luminosity in order to improve the Higgs couplings and production cross section measurements. Such a precise measurement at these Higgs factories, where the uncertainty would be much smaller, will be important to discover the evidences of new physics beyond the standard model. For all these reasons, we have investigated the laser-assisted neutral Higgs-boson pair production process in IHDM in order to study the effect of the electromagnetic field on its cross section.

The remainder of this research paper is organized as follows: The section 2 is devoted to the theoretical calculation of the total cross section of the neutral Higgs pair production process (${e}^{+}{e}^{-}\rightarrow H^{0}A^{0}$) in the presence of a circularly polarized electromagnetic field. The results obtained are discussed in section 3. A short conclusion is given in section 4. In the appendix, we have listed the some of the multiplying coefficients of the Bessel function that appears in equation \ref{18}. In this work, we have adopted natural units such that $(\hbar = c = 1)$. The Livi-Civita tensor is chosen such that $\varepsilon^{0123}=1$, and the metric $g^{\mu\nu}$ is taken as $g^{\mu\nu}=(1,-1,-1,-1)$.  
\section{Outline of the theory}\label{sec:theory}
\begin{figure}[H]
  \centering
      \includegraphics[width=0.5\textwidth]{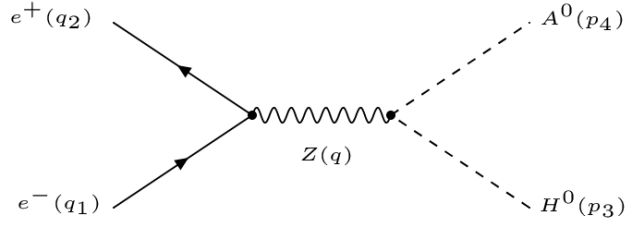}
  \caption{Feynman diagram  for $s$-channel neutral higgs-boson pair production in the lowest order.}
  \label{fig1}
\end{figure}

This part is devoted to the theoretical calculation of the differential cross section for the process of neutral Higgs-boson pair production ${e}^{+}{e}^{-}\rightarrow H^{0}A^{0}$, at the lowest order, which is described by the Feynman diagram as in figure \ref{fig1}.
The scattering matrix element \cite{25} of this process can be expressed as follows:
\small
\begin{eqnarray}
S_{fi}({e}^{+}{e}^{-}\rightarrow H^{0}A^{0})&=&\frac{-ie}{2C_{W} S_{W} }  \int_{}^{} d^4x \int d^4y  \bar{\psi}_{p_{2},s_{2}}(x)   \Big[ \gamma^{\alpha} (g_v^{e} -g_a^{e}\gamma^{5}) \Big]   \psi_{p_{1},s_{1}}(x)  \nonumber \\
 &\times & D_{\alpha \sigma}(x-y)  \phi^{*}_{p_{3}}(y)   \Big( \dfrac{e \overleftrightarrow{\partial_{\alpha}}}{2C_{W} S_{W}} \Big)  \phi^{*}_{p_{4}}(y),
 \label{1}
\end{eqnarray}
\normalsize
 where $C_{W}=\cos_{\theta_{W}}$, and $S_{W}=\sin_{\theta_{W}}$, with $\theta_W$ is the Weinberg angle. $e$ is the electron charge. $x$ is the space time coordinate of the electron and positron, and $y$ is the space time coordinate of the scattered Higgs-bosons. $g_v^{e}$ is the vector coupling constant, and $g_a^{e}$ is the axial vector coupling constant. The matrix $\gamma^{5}$ is expressed in terms of Dirac matrix as follows: $\gamma^{5}=\gamma^{0}\gamma^{1}\gamma^{2}\gamma^{3}\gamma^{4}$. The derivative $\overleftrightarrow{\partial_{\alpha}}$ is defined such that: $f\overleftrightarrow{\partial}g=f*\partial g -\partial f*g$, where $f$ and $g$ are a given functions. The $Z^{*}$-boson propagator $ D_{\alpha \sigma}(x-y) $  \cite{25} is given by:
\begin{equation}
D_{\alpha\sigma}(x-y)=\int \dfrac{d^{4}q}{(2\pi)^4} \frac{e^{-iq(x-y)}}{q^{2}-M_{Z}^{2}}\Big[-ig_{\alpha\sigma}+i\dfrac{q_{\alpha}q_{\sigma}}{M_{Z}^{2}}\Big],
\label{2}
\end{equation}
where $q$ is the four-momentum of the $Z^{*}$-boson propagator.
$\psi_{p_{1},s_{1}}(x)$ is the Dirac-Volkov state \cite{26} of the electron, and $\psi_{p_{2},s_{2}}(x)$ is the Dirac-Volkov state of the positron inside the electromagnetic field. They are given by:
\begin{equation}
\begin{cases}
\psi_{p_{1},s_{1}}(x)= \Big[1-\dfrac{e \slashed k \slashed A}{2(k.p_{1})}\Big] \frac{u(p_{1},s_{1})}{\sqrt{2Q_{1}V}} \exp^{iS(q_{1},s_{1})}&\\
\psi_{p_{2},s_{2}}(x)= \Big[1+\dfrac{e \slashed k \slashed A}{2(k.p_{2})}\Big] \frac{v(p_{2},s_{2})}{\sqrt{2Q_{2}V}} \exp^{iS(q_{2},s_{2})},
\end{cases}
\label{3}
\end{equation}
with, $u(p_{1},s_{1})$ is the bispinor of the electron, and $v(p_{2},s_{2})$ is the bispinor of the positron. $p_{i}(i=1,2)$ is the free momentum of the electron and positron, and $s_{i}(i=1,2)$ is its corresponding spin such that: $\sum_{s}^{}u(p_{1},s_{1})\bar{u}(p_{1},s_{1})=(\slashed p_{1}-m_{e})$ and $\sum_{s}^{}v(p_{2},s_{2})\bar{v}(p_{2},s_{2})=(\slashed p_{2}+m_{e})$.
$q_{i}=(Q_{i},q_{i})(i=1,2)$  is the effective four-momentum of the electron and the positron, and $ Q_{i}(i=1,2)$ is the effective energy acquired by the incident particles. The effective four-momentum and the laser-free four-momentum are related by the following equation $q_{i}=p_{i}+(e^{2}a^{2}/2(k.p_{i}))k$. From this equation we can derive the square of the effective momentum such that: $ q_{i}^{2}=m_{e}^{*^{2}}=(m_{e}^{2}+e^{2}a^{2})$, where $m_{e}^{*}$ is the effective mass of the electron and positron. $\slashed A=\gamma_{\alpha}A^{\alpha}$ with $A^{\alpha}$ is the classical four-potential  which is defined as follows:
\begin{equation}
A^{\alpha}(\phi)=a_{1}^{\alpha}\cos\phi+a_{2}^{\alpha}\sin\phi \hspace*{1cm};\hspace*{1cm} \phi=(k.x)
\label{4}
\end{equation}
where $a_{1,2}^{\alpha}$ are the polarization four vectors which are chosen as  $a_{1}^{\alpha}=(0,a,0,0)$ and $a_{2}^{\alpha}=(0,0,a,0)$, where $a$ denotes the amplitude of the four-potential such that $(a_{1}.a_{2})=0$, and $a_{1}^{2}=a_{2}^{2}=a^{2}=-|\mathbf{a}|^{2}=-\big(\epsilon_{0}^{2}/\omega \big)$, with $\epsilon_{0}$ is the laser field strength.
The Lorentz gauge is satisfied $\partial_{\alpha}A^{\alpha}=0$, which implies that $(k_{\alpha}A^{\alpha}=0;\, a_{1}.k=0;\, a_{2}.k=0)$, where the vector $k$ is chosen as parallel to the $z$-axis such that $k=(\omega,0,0,\omega)$ and $k^{2}=0)$. $\phi$ is the phase of the laser field, and $\omega$ its frequency.
The arguments of the exponential terms in equation \ref{3} are defined as follows:
\begin{equation}
\begin{cases}
S(q_{1},s_{1})=- q_{1}x +\frac{e(a_{1}.p_{1})}{k.p_{1}}\sin\phi - \frac{e(a_{2}.p_{1})}{k.p_{1}}\cos\phi &\\
S(q_{2},s_{2})=+ q_{2}x +\frac{e(a_{1}.p_{2})}{k.p_{2}}\sin\phi - \frac{e(a_{2}.p_{2})}{k.p_{2}}\cos\phi
\end{cases}
\label{5}
\end{equation}
The scattered particles are neutral Higgs-bosons with spin 1. Therefore, they do not interact with the electromagnetic field, and can be described by using the Klein-Gordon state as follows:
\begin{equation}
\phi_{p_{3}}(y)=\dfrac{1}{\sqrt{2 Q_{H^{0}} V}} e^{-ip_{3}y} \hspace*{0.5cm};\hspace*{0.5cm}  \\\ \phi_{p_{4}}(y)=\dfrac{1}{\sqrt{2 Q_{A^{0}} V}} e^{-ip_{4}y},
\label{6}
\end{equation}
where $p_{3}$ is the four-momenta of CP-even Higgs-boson $H^{0}$, and $p_{4}$ is the four-momenta of the CP-odd Higgs-boson $A^{0}$. $Q_{H^{0}}$ and $Q_{A^{0}}$ are its corresponding energies. After some algebraic calculation, the scattering matrix element will be as in the following equation: 
\begin{eqnarray}
S_{fi}^{n}({e}^{+}{e}^{-}\rightarrow H^{0}A^{0})&=&\dfrac{1}{4V^{2}\sqrt{Q_{1}Q_{2}Q_{A^{0}}Q_{H^{0}}}}\dfrac{e^{2}}{4C_{W}^{2} S_{W}^{2}}  \Bigg( \dfrac{1}{(q_{1}+q_{2}+nk)^{2}-M_{Z}^{2}}  \Bigg) \nonumber \\
 &\times & (p_{3}^{\alpha}-p_{4}^{\alpha}) \bar{v}(p_{2},s_{2})\Gamma^{n}_{\alpha}u(p_{1},s_{1}),
 \label{7}
\end{eqnarray}
where $n$ is the number of exchanged photons between the laser field and the colliding physical system. It can be either negative (emission of photons) or positive (absorption of photons). The quantity $\Gamma^{n}_{\alpha}$ is expressed as follows:
\begin{equation}
 \Gamma_{\alpha}^{n}=B^{0}_{\alpha}\,D_{0n}(z)+B^{1}_{\alpha}\,D{1n}(z)+B^{2}_{\alpha}\,D_{2n}(z),
 \label{8}
 \end{equation}
where the coefficients $D_{0n}(z)$, $D_{1n}(z)$ and $D_{2n}(z)$ can be expressed as follows:
\begin{equation}
\left.
  \begin{cases}
     D_{0n}(z) \\
      D_{1n}(z) \\
      D_{2n}(z)
  \end{cases}
  \right\} = \left.
  \begin{cases}
     J_{n}(z)e^{-in\phi _{0}}\\
    \frac{1}{2}\Big(J_{n+1}(z)e^{-i(n+1)\phi _{0}}+J_{n-1}(z)e^{-i(n-1)\phi _{0}}\Big) \\
     \frac{1}{2\, i}\Big(J_{n+1}(z)e^{-i(n+1)\phi _{0}}-J_{n-1}(z)e^{-i(n-1)\phi _{0}}\Big)
  \end{cases}
  \right\}.
  \label{9}
\end{equation}
The argument of the Bessel function $z$ and the phase $\phi_{0}$ are given by:
$ z=\sqrt{\beta_{1}^{2}+\beta_{2}^{2}}$ and $\phi_{0}= \arctan(\beta_{2}/\beta_{1})$, where:
\begin{center}
$\beta_{1}=\dfrac{e(a_{1}.p_{1})}{(k.p_{1})}-\dfrac{e(a_{1}.p_{2})}{(k.p_{2})}$ \qquad \qquad $\beta_{2}=\dfrac{e(a_{2}.p_{1})}{(k.p_{1})}-\dfrac{e(a_{2}.p_{2})}{(k.p_{2})}$.\\
\end{center}
The quantities $B^{0}_{\alpha}$, $B^{1}_{\alpha}$ and $B^{2}_{\alpha}$ that appear in equation \ref{8} are given by the following expressions:
\begin{equation}
\begin{cases}B^{0}_{\alpha}=\gamma_{\alpha}(g_{v}^{e}-g_{a}^{e}\gamma^{5})+2b_{p_{1}}b_{p_{2}}a^{2}k_{\alpha}\slashed k(g_{v}^{e}-g_{a}^{e}\gamma^{5})   &\\
B^{1}_{\alpha}=b_{p_{1}}\gamma_{\alpha}(g_{v}^{e}-g_{a}^{e}\gamma^{5})\slashed k\slashed a_{1}-b_{p_{2}}\slashed a_{1}\slashed k \gamma_{\alpha}(g_{v}^{e}-g_{a}^{e}\gamma^{5})   &\\
B^{2}_{\alpha}=b_{p_{1}}\gamma_{\alpha}(g_{v}^{e}-g_{a}^{e}\gamma^{5})\slashed k\slashed a_{2}-b_{p_{2}}\slashed a_{2}\slashed k \gamma_{\alpha}(g_{v}^{e}-g_{a}^{e}\gamma^{5}) \end{cases},
\label{10}
\end{equation}
with $b_{p_{i}}=e/2(kp_{i})$. The differential cross section, in the center of mass frame, is derived by squaring and dividing the scattering matrix element by $VT$ to get the transition probability per volume, by $|J_{inc}|=(\sqrt{(q_{1}q_{2})^{2}-m_{e}^{*^{4}}}/{Q_{1}Q_{2}V})$, and by the density of the particles $\rho=V^{-1}$. Finally, since the differential cross section is unpolarized, we have to sum over the final spins and average over the initial ones. Therefore, we obtain:
\small
\begin{eqnarray}
d\sigma_{n}&=&\dfrac{e^{4}}{256C_{W}^{4} S_{W}^{4}}  \Bigg[ \dfrac{1}{(q_{1}+q_{2}+nk)^{2}-M_{Z}^{2}}  \Bigg]^{2}  \dfrac{1}{\sqrt{(q_{1}q_{2})^{2}-m_{e}^{*^{4}}}} \big|\overline{M_{fi}^{n}} \big|^{2}   \int \dfrac{2|\mathbf{p}_{3}|^{2}d|\mathbf{p}_{3}|d\Omega}{(2\pi)^{2}Q_{H^{0}}}   \nonumber \\
 &\times &  \int \dfrac{d^{3}p_{4}}{Q_{A^{0}}} \delta^{4}(p_{3}+p_{4}-q_{1}-q_{2}-nk).
\label{11}
\end{eqnarray}
\normalsize
Now, we have to integrate over $d^{3}p_{4}$, and use the well known formula given by:
\begin{equation}
 \int d\mathbf y f(\mathbf y) \delta(g(\mathbf y))=\dfrac{f(\mathbf y)}{|g^{'}(\mathbf y)|_{g(\mathbf y)=0}}
 \label{12}
\end{equation}
Thus, the differential cross section becomes as expressed in the following equation:
\begin{eqnarray}
\dfrac{d\sigma_{n}}{d\Omega}&=&\dfrac{e^{4}}{256C_{W}^{4} S_{W}^{4}}  \Bigg[ \dfrac{1}{(q_{1}+q_{2}+nk)^{2}-M_{Z}^{2}}  \Bigg]^{2}  \dfrac{1}{\sqrt{(q_{1}q_{2})^{2}-m_{e}^{*^{4}}}}  \big|\overline{M_{fi}^{n}} \big|^{2}   \dfrac{2|\mathbf{p_{3}}|^{2}}{(2\pi)^{2}Q_{H^{0}}}\nonumber \\
 &\times & \dfrac{1}{\big|g^{'}(|\mathbf{p}_{3}|)\big|_{g(|\mathbf{p}_{3}|)=0}},
\label{13}
\end{eqnarray}
where $g^{'}(|\mathbf{p}_{3}|)$ is given by:
\begin{equation}
 g^{'}(|\mathbf{p}_{3}|)=\dfrac{4 e^{2}a^{2}}{\sqrt{s}}\dfrac{|\mathbf{p}_{3}|}{\sqrt{|\mathbf{p}_{3}|^{2}+M_{H^{0}}^{2}}}-\dfrac{2|\mathbf{p}_{3}|(\sqrt{s}+n\omega)}{\sqrt{|\mathbf{p}_{3}|^{2}+M_{H^{0}}^{2}}}
 \label{14}
\end{equation}
The term $ \big|\overline{M_{fi}^{n}} \big|^{2}$ that appears in equation \ref{11} can be evaluated as follows:
\begin{equation}
\big|\overline{M_{fi}^{n}} \big|^{2}=\dfrac{1}{4}\sum_{n=-\infty}^{+\infty}\sum_{s}\big|M_{fi}^{n} \big|^{2}=\frac{1}{4}\sum_{n=-\infty}^{+\infty}Tr\Big[ (\slashed p_{2}+m_{e})  (p_{3}^{\alpha}-p_{4}^{\alpha})    \Gamma^{n}_{\alpha}(\slashed p_{1}-m_{e})(p_{3}^{\sigma}-p_{4}^{\sigma}) \bar{\Gamma}^{n}_{\sigma}\Big],
\label{15}
\end{equation}
where $\Gamma^{n}_{\alpha}$ is given by equation \ref{8}, and $\bar{\Gamma}^{n}_{\sigma}$ is expressed as follows:
\begin{equation}
\bar{\Gamma}_{\sigma}^{n}=\bar{B^{0}_{\sigma}}\,D_{0n}^{*}(z)+\bar{B^{1}_{\sigma}}\,D_{1n}^{*}(z)+\bar{B^{2}_{\sigma}}\,D_{2n}^{*}(z),
\label{16}
 \end{equation}
with: 
\small
\begin{equation}
\begin{cases}\bar{B^{0}_{\sigma}}=\gamma_{\sigma}(g_{v}^{e}-g_{a}^{e}\gamma^{5})+2b_{p_{1}}b_{p_{2}}a^{2}k_{\sigma}\slashed k(g_{v}^{e}-g_{a}^{e}\gamma^{5})   &\\
\bar{B^{1}_{\sigma}}=b_{p_{1}} \slashed a_{1} \slashed k \gamma_{\sigma}(g_{v}^{e}-g_{a}^{e}\gamma^{5}) -b_{p_{2}}\gamma_{\sigma}(g_{v}^{e}-g_{a}^{e}\gamma^{5})\slashed k\slashed a_{1}   &\\
\bar{B^{2}_{\sigma}}=b_{p_{1}} \slashed a_{2} \slashed k \gamma_{\sigma}(g_{v}^{e}-g_{a}^{e}\gamma^{5}) -b_{p_{2}}\gamma_{\sigma}(g_{v}^{e}-g_{a}^{e}\gamma^{5})\slashed k\slashed a_{2} \end{cases}.
\label{17}
\end{equation}
\normalsize
The trace in equation \ref{15} is numerically calculated by using FeynCalc program \cite{27}, and the result obtained has the following form:
 \small
\begin{eqnarray}
\big|\overline{M_{fi}^{n}} \big|^{2}&=&\frac{1}{4}\sum_{n=-\infty}^{+\infty}\Big[ AJ_{n}^{2}(z)+BJ_{n+1}^{2}(z)+CJ_{n-1}^{2}(z)+DJ_{n}(z)J_{n+1}(z)+EJ_{n}(z)J_{n-1}(z)\nonumber \\
 &+ & FJ_{n-1}(z)J_{n+1}(z) \Big].
\label{18}
\end{eqnarray}
\normalsize
By using FeynCalc tools, we have derived the expressions of the coefficients $A$, $B$, $C$, $D$, $E$ and $F$. We give here the expression of the first coefficient multiplied by $J_{n}^{2}(z)$, the others are given in the appendix.
\small
\begin{eqnarray}
A&=&\nonumber\dfrac{1}{(k.p_{1})(k.p_{2})}\Big[ 2 (a^4 e^4 (g_{a}^{e^{2}} + g_{v}^{e^{2}}) ((k.p_{3}) - (k.p_{4}))^2 + 
    2 a^2 e^2 ((k.p_{3}) - (k.p_{4})) (g_{a}^{e^{2}} (((k.p_{3}) - (k.p_{4}))\\&\times &\nonumber (m_{e}^{2} - (p_{1}.p_{2})) + (k.p_{2}) (p_{1}.p_{3}) - (k.p_{2}) (p_{1}.p_{4}) + 
          (k.p_{1}) (p_{2}.p_{3}) - (k.p_{1}) (p_{2}.p_{4})) + g_{v}^{e^{2}} (-((k.p_{3}) \\&- &\nonumber (k.p_{4})) (m_{e}^{2} + (p_{1}.p_{2})) + (k.p_{2}) (p_{1}.p_{3}) - (k.p_{2}) (p_{1}.p_{4}) + (k.p_{1}) (p_{2}.p_{3}) - (k.p_{1}) (p_{2}.p_{4}))) \\&+ &\nonumber
    2 (k.p_{1}) (k.p_{2}) (2 g_{a}^{e^{2}} (((p_{1}.p_{3}) - (p_{1}.p_{4})) ((p_{2}.p_{3}) - (p_{2}.p_{4})) + (m_{e}^{2} - 
             (p_{1}.p_{2})) (m_{A^{0}}^{2} - (p_{3}.p_{4}))) \\&- &\nonumber 
       g_{v}^{e^{2}} (m_{e}^{2} m_{H^{0}}^{2} + m_{A^{0}}^{2} (m_{e}^{2} + 2 (p_{1}.p_{2})) - 
          2 ((p_{1}.p_{3}) - (p_{1}.p_{4})) ((p_{2}.p_{3}) - (p_{2}.p_{4})) \\&- & 2 (m_{e}^{2} + (p_{1}.p_{2})) (p_{3}.p_{4}))))\Big].
\label{19}
\end{eqnarray}
\normalsize
 The total cross section is obtained by numerically integrating over the solid angle $d\Omega=\sin(\theta)d\theta d\phi$, where $\theta$ is the scattering angle.
\section{Results and Discussion}\label{sec:results}
In this research paper's part, we will discuss, in the centre of mass frame, the numerical results about the behavior of the laser-assisted total cross section of the process ${e}^{+}{e}^{-}\rightarrow H^{0}A^{0}$. This cross section depends on the centre of mass energy, the mass of $H^{0}$, the mass of $A^{0}$ and the laser field parameters. Every $\sigma_{n}$, considering four-momentum conservation $p_{3}+p_{4}-q_{1}-q_{2}-nk$, can be interpreted as the partial total cross section that describes the scattering process, and by summing over a number of exchanged photons, we obtain the total cross section. We will begin our discussion by showing how the partial total cross section varies as a function of the photons number $n$. Then, we show how it depends on the outgoing particles mass for different number of exchanged photons. Next, we illustrate its variation as a function of the centre of mass energy for different neutral Higgs-boson masses. Finally, we indicate how it changes as a function of the laser field amplitudes for both different number of exchanged photons and different Higgs-boson masses. We should mention that, throughout this work, the outgoing neutral Higgs-bosons are considered to be degenerate, i.e. they have the same mass such that: $m_{A^{0}}=m_{H^{0}}=m_{s}$. In addition, we have taken the Higgs-boson mass as $m_{s}=120\,GeV$ in all figures except figure \ref{fig4}. We have also adopted the following numerical values from PDG \cite{28}: $m_{Z}=91.1875\,GeV$ and $m_{e}=0.511\,MeV$.
\begin{figure}[H]
  \centering
      \includegraphics[scale=0.58]{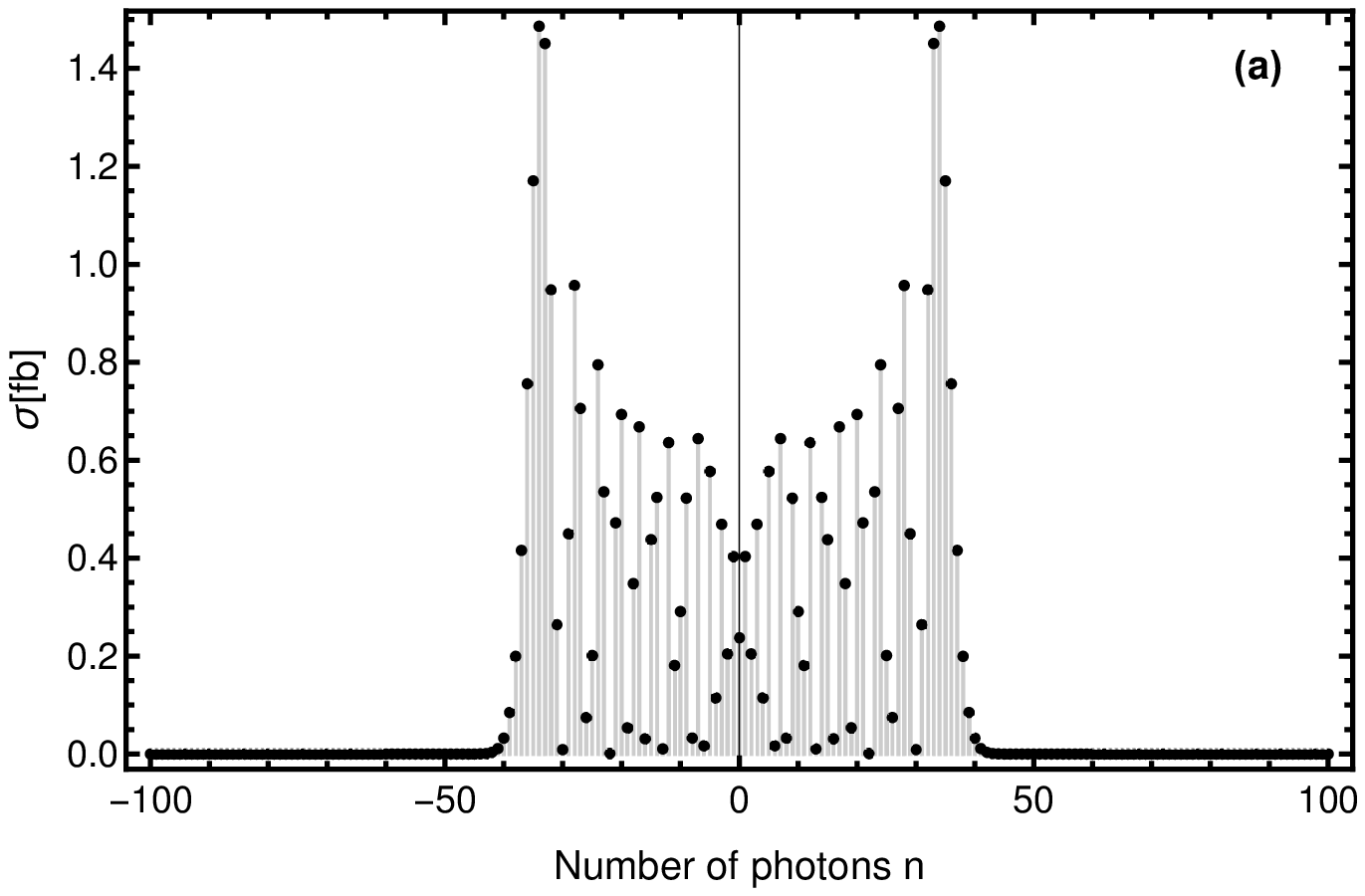}\hspace*{0.4cm}
      \includegraphics[scale=0.58]{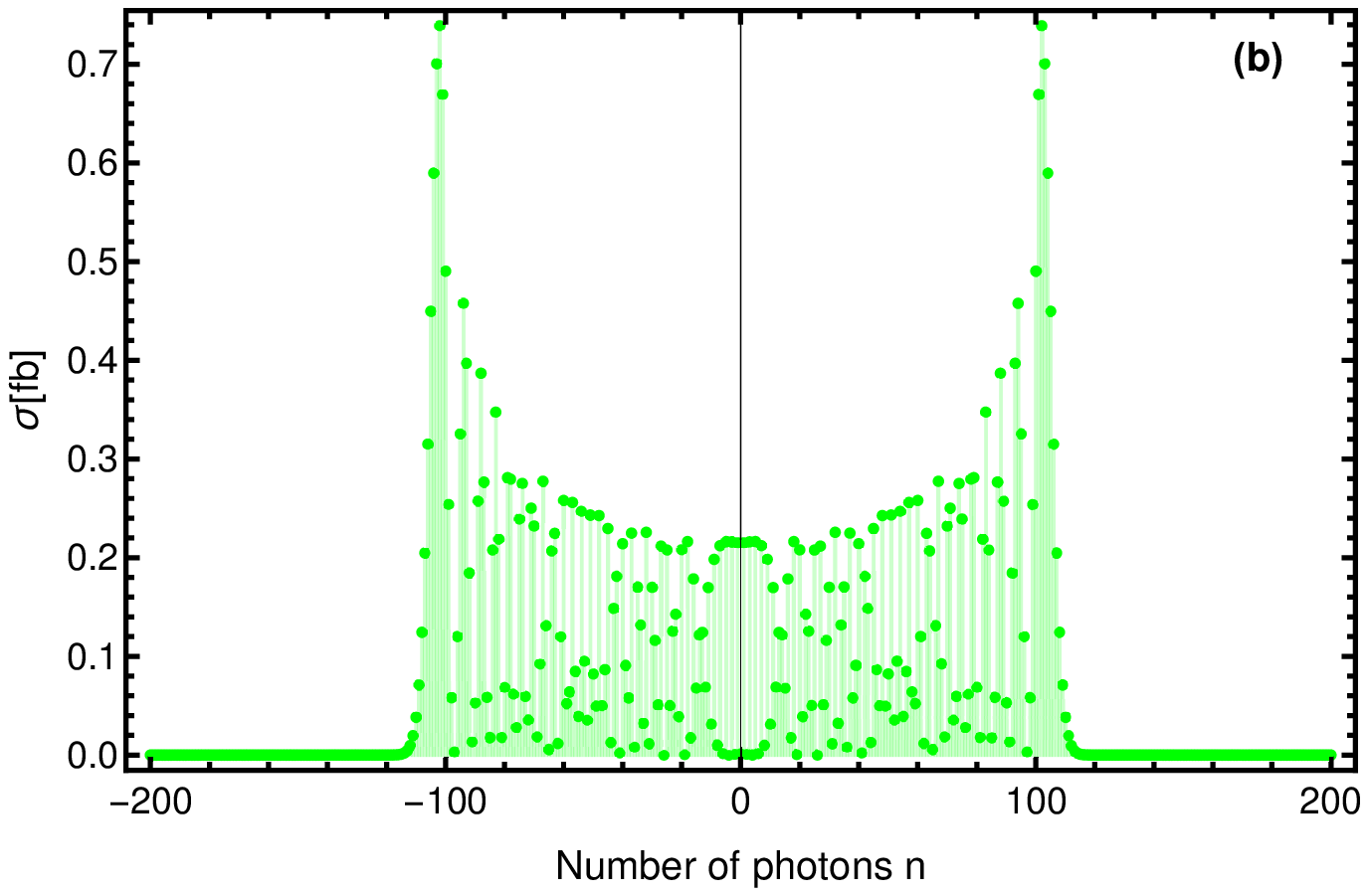}\par\vspace*{0.5cm}
      \includegraphics[scale=0.58]{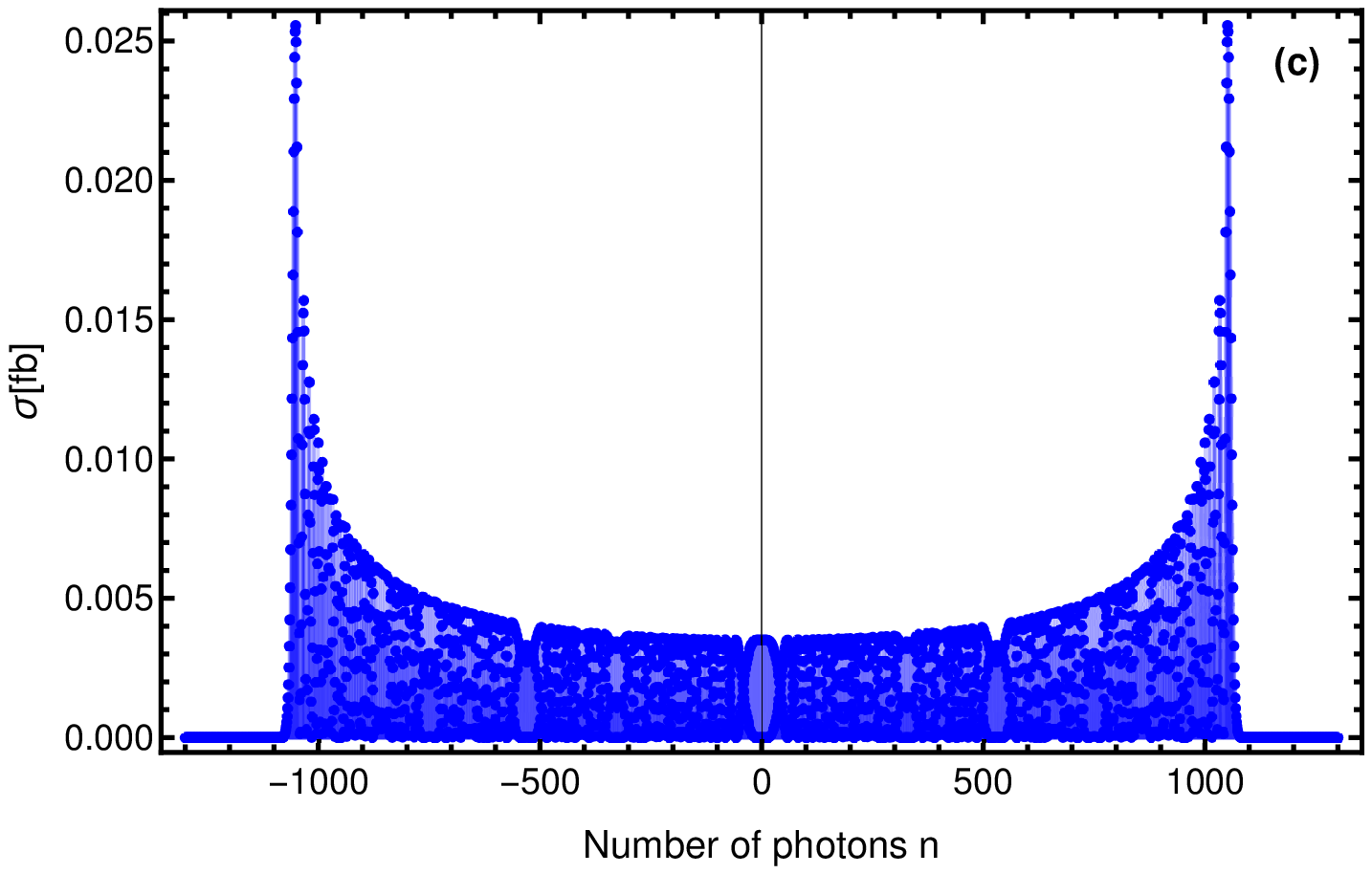}
        \caption{Dependence of the laser-assisted partial total cross section of the neutral Higgs-boson pair production on the number of exchanged photons. The laser field strength and its frequency are taken as: $\epsilon_{0}=10^{6}V.cm^{-1}$ and $\omega=2\, eV$ in (a);  $\epsilon_{0}=10^{6}V.cm^{-1}$ and $\omega=1.17\, eV$ in (b);  $\epsilon_{0}=10^{7}V.cm^{-1}$ and $\omega=1.17\, eV$ in (c).} 
        \label{fig2}
\end{figure}
Figure \ref{fig2} displays the dependence of the partial total cross section of the process ${e}^{+}{e}^{-}\rightarrow H^{0}A^{0}$ on the number of exchanged photons for different laser field strengths and frequencies. This dependence gives us much information about the required number of photons to reach the well known sum-rule, which states that the summation of the partial total cross section over the cutoff number converges to the laser-free total cross section. This sum-rule was first elaborated by Kroll and Watson \cite{29}, and it is given by the following equation:
\begin{equation}
\sum_{n=-cutoff}^{n=+cutoff} \sigma_{n}=\sigma
\end{equation} 
 All curves present cutoffs at two edges which are symmetric with respect to $n=0$, where $n>0$ corresponds to absorption of photons while $n<0$ corresponds to the emission. In addition, these cutoffs varies form one figure to another as they depends on the laser field parameters, i.e. laser field amplitude and its frequency. By comparing figures (\ref{fig2}a) and (\ref{fig2}b), we observe that, for the same laser field amplitude which is $\epsilon_{0}=10^{6}V.cm^{-1}$, the cutoff number increases as far as the laser frequency decreases. In contrast, by fixing the laser frequency as in figures (\ref{fig2}b) and (\ref{fig2}c), we can conclude that the cutoff number increases as long as the laser field amplitude increases. 
To understand clearly why we have involved the sum-rule, let's analyze how the total cross section of this scattering process varies as a function of the outgoing neutral Higgs-boson mass for different number of exchanged photons.
\begin{figure}[H]
  \centering
      \includegraphics[scale=0.58]{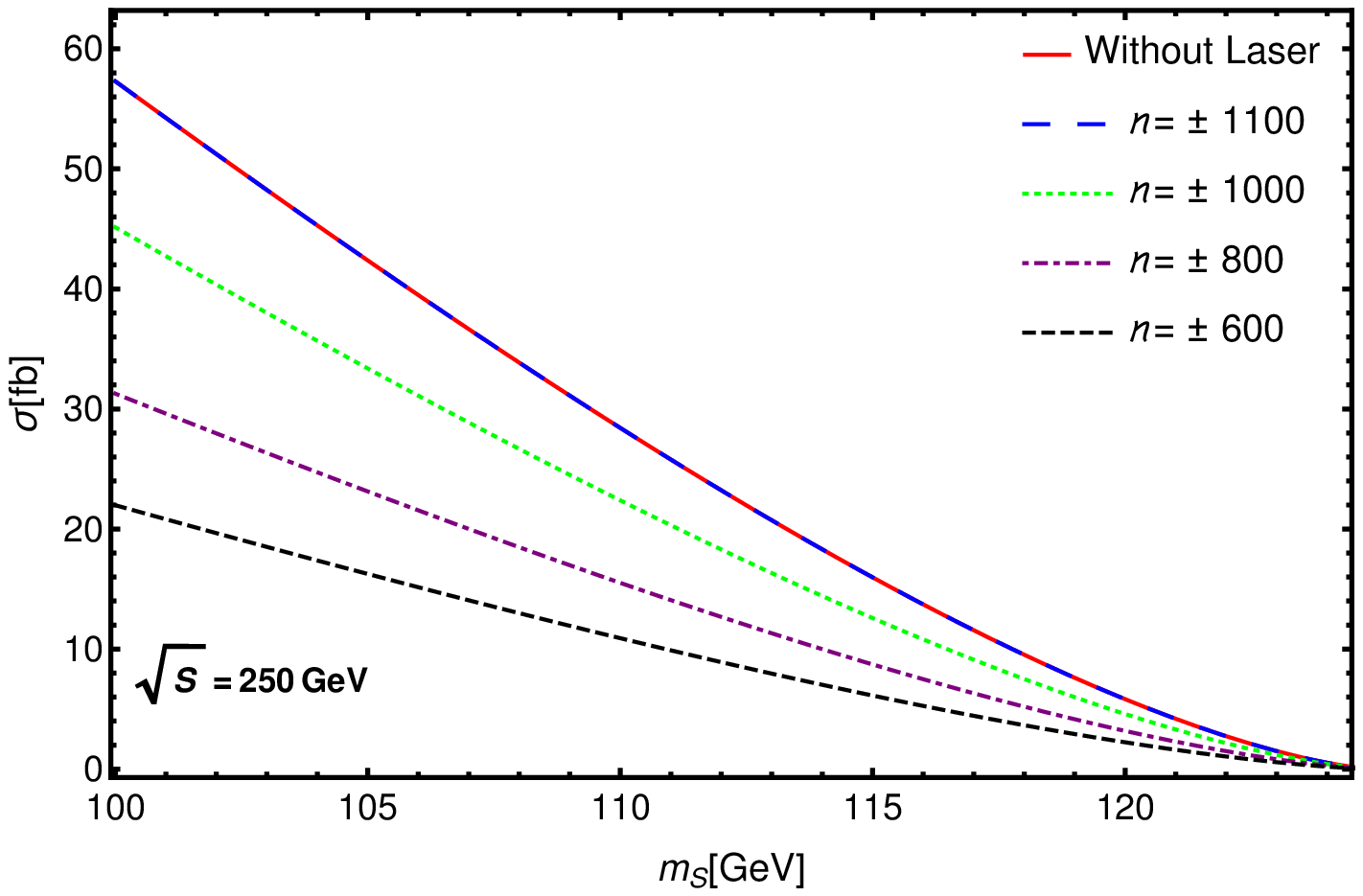}\hspace*{0.4cm}
      \includegraphics[scale=0.58]{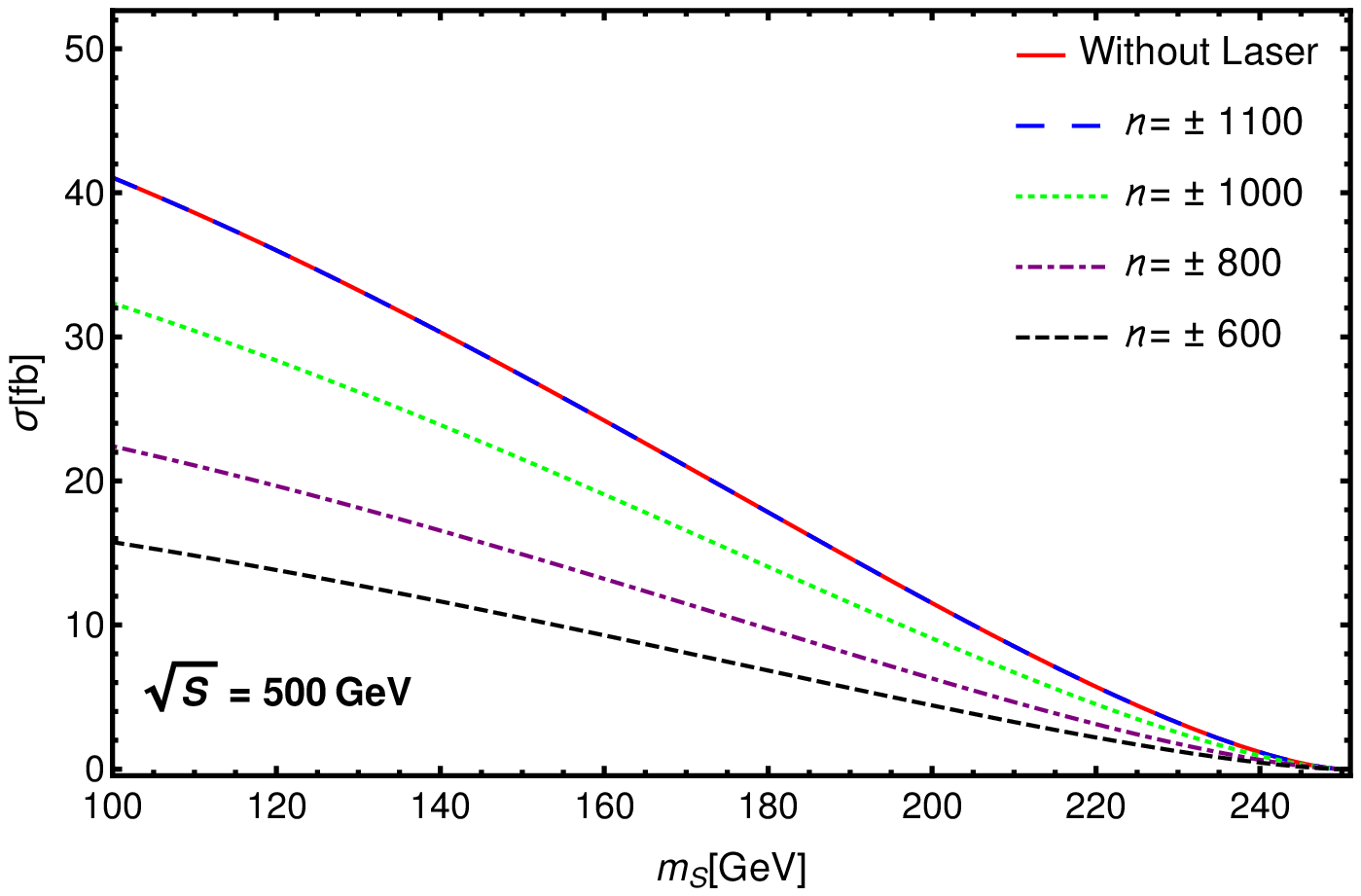}\par\vspace*{0.5cm}
      \includegraphics[scale=0.58]{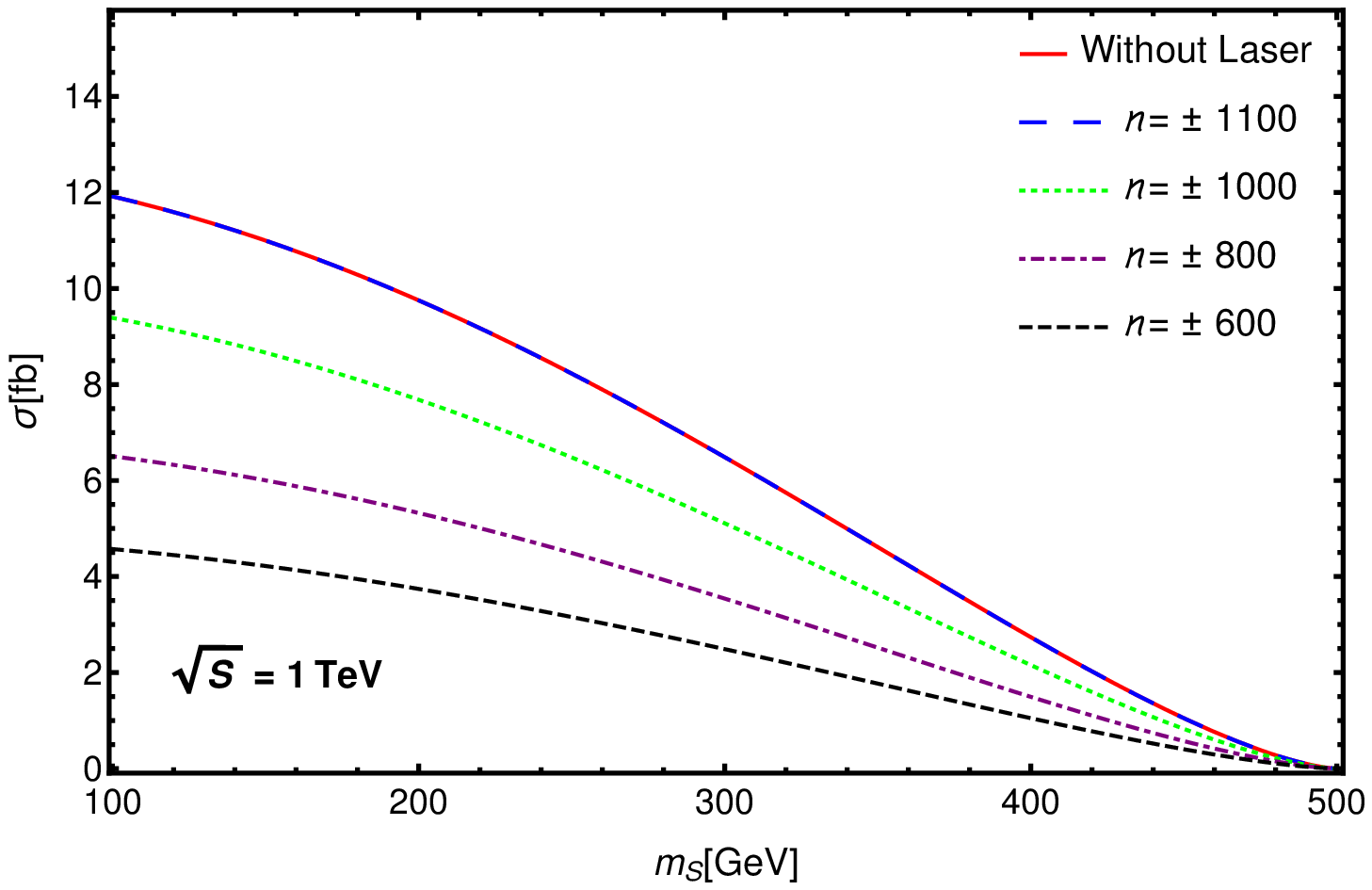}
        \caption{Dependence of the laser-assisted total cross section of the process ${e}^{+}{e}^{-}\rightarrow H^{0}A^{0}$ on the outgoing Higgs-boson mass for different number of exchanged photons and with three centre of mass energies: $250GeV$, $500GeV$ and $1TeV$.}
        \label{fig3}
\end{figure}
Figure \ref{fig3} shows the laser-assisted total cross section of the neutral Higgs-boson pair production in the IHDM for three typical centre of mass energies which are $250\,GeV$, $500\,GeV$ and $1\,TeV$. The laser field strength and its frequency are chosen as $\epsilon_{0}=10^{7}V.cm^{-1}$ and $\omega=1.17\, eV$, respectively. It is obvious that, regardless of the centre of mass energy value, the laser-free total cross sections (red curves) decrease as far as the neutral Higgs-boson mass increases. However, its order of magnitudes decrease as much as the centre of mass energy increases. For example, for $m_{s}=100\,GeV$, the value of the total cross section is $\sigma=45[fb]$ for $\sqrt{s}=500\,GeV$ while its value for $\sqrt{s}=500\,GeV$ is equal to $\sigma=12[fb]$. By applying the electromagnetic field and for all the centre of mass energies, we observe that the total cross section decreases by several orders of magnitudes. In addition, this decreasing process depends on the number of exchanged photons between the laser field and the physical system. Moreover, by increasing the number of exchanged photons, we notice that the laser-assisted total cross section converges to the laser-free total cross section. Furthermore, if we perform the summation over the number of exchanged photons from $-1100$ to $+1100$, the laser-assisted total cross section will be equal to its corresponding laser-free total cross section. This result is in full agreement with our previous work \cite{11}, in which we have shown the effect of a circularly polarized laser field on the Higgs-strahlung cross section. It is interpreted by the fact that the available phase space is reduced inside the electromagnetic field. Consequently, the cross-section obviously falls down. Since the neutral Higgs-bosons are not experimentally discovered yet, it is of great importance, therefore, to study the behavior of the total cross section of the process
$({e}^{+}{e}^{-}\rightarrow H^{0}A^{0}$) as a function of the Higgs-boson mass.
\begin{figure}[H]
  \centering
      \includegraphics[scale=0.80]{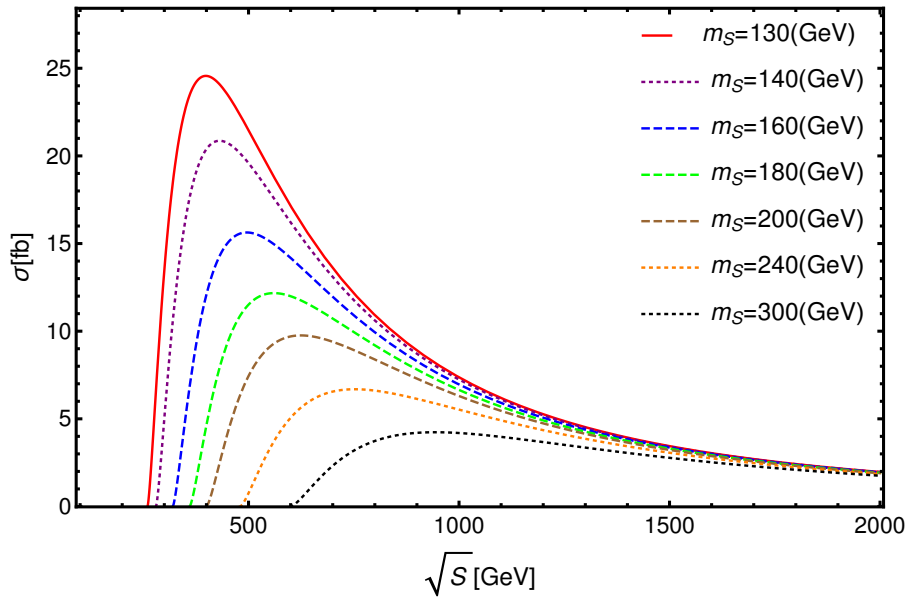}\hspace*{0.4cm}
        \caption{Variation of the laser-assisted total cross section of the process ${e}^{+}{e}^{-}\rightarrow H^{0}A^{0}$ as a function of the centre of mass energy for different neutral Higgs-boson masses by summing over $n$ from -900 to +900 and taking the laser's parameters such as: $\epsilon_{0}=10^{7}V.cm^{-1}$, $\omega=1.17\, eV$.}
        \label{fig4}
\end{figure}
Figure \ref{fig4} displays the dependence of the laser-assisted total cross section of the process ${e}^{+}{e}^{-}\rightarrow H^{0}A^{0}$ as a function of the centre of mass energy for different neutral Higgs-boson masses and by taking the laser field strength and frequency as:  $\epsilon_{0}=10^{7}V.cm^{-1}$ and $\omega=1.17\, eV$. It is clear that, for the range of low centre of mass energies (e.g. $\sqrt{s} \leq 250\,GeV$ for $m_{s}=130\,GeV$), there is no probability for producing the pair of neutral Higgs-bosons. However, the threshold value of the centre of mass energy from which the total cross section begins to raise increases as far the Higgs-boson mass increases. Whereas, whenever the centre of mass energy overcomes this threshold value and regardless of the Higgs-boson mass, the total cross section increases rapidly until it reaches its maximum value, which varies from one Higgs-boson mass to another. Then it begins to decrease slowly by raising the centre of mass energy. By comparing The maximum values of the total cross section and its corresponding centre of mass energies, we remark that they depend on the neutral Higgs-boson mass. To be more precise, the high total cross section maximum value occurs at low centre of mass energies for low Higgs-boson masses. Moreover, this maximum value decreases by increasing the Higgs-boson mass. Furthermore, high Higgs-boson masses require high centre of mass energies for the process to reach the maximum of the cross section. For instance, for $m_{s}=130\,GeV$, $\sigma_{max}=24.54\,GeV$, and it occurs at $\sqrt{s}=396\,GeV$. However, for $m_{s}=300\,GeV$, $\sigma_{max}=4.312\,GeV$, and it occurs at $\sqrt{s}=911\,GeV$. No, let's focus our attention on how the laser-assisted total cross section depends on the laser field parameters, i.e. the laser field strength, its frequency and the number of exchanged photons.
\begin{figure}[H]
  \centering
      \includegraphics[scale=0.60]{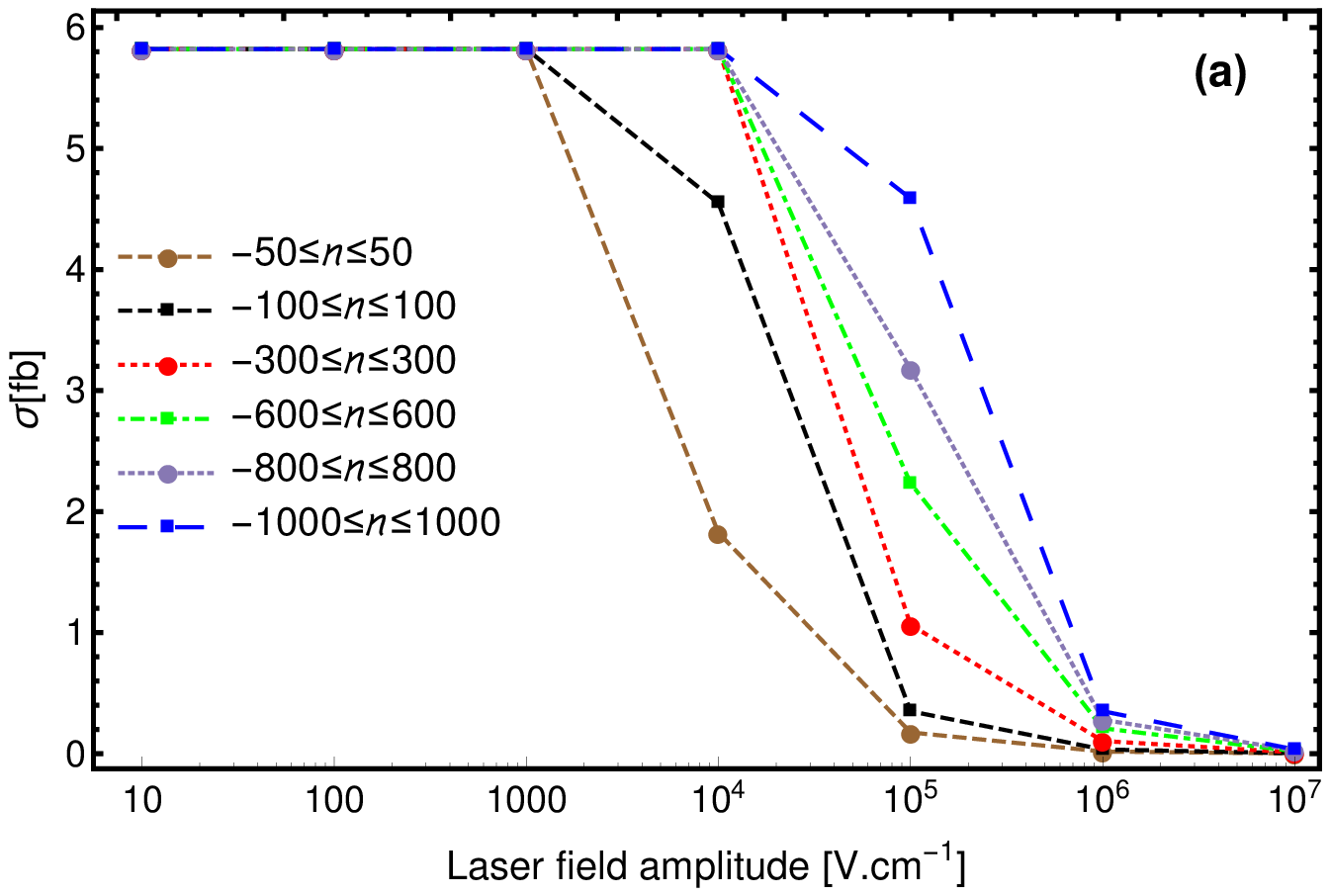}\hspace*{0.4cm}
      \includegraphics[scale=0.60]{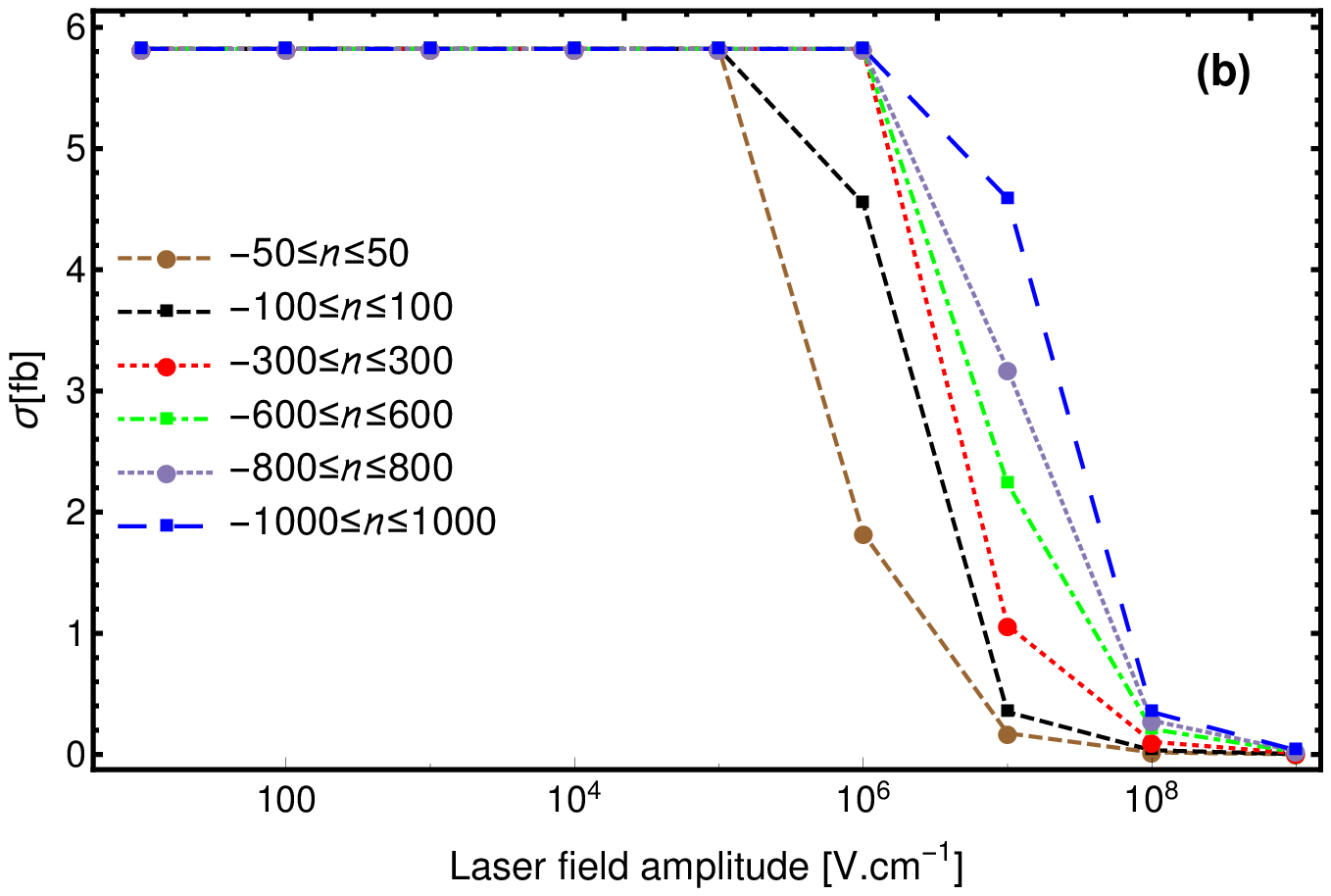}\par\vspace*{0.5cm}
        \caption{Laser-assisted total cross section of the process ${e}^{+}{e}^{-}\rightarrow H^{0}A^{0}$ as a function of the laser field strength for different exchanged photons number. The laser frequency  is taken as  $\,\omega=0.117 eV$ in (a) and $\omega=1.17\,eV$ in (b).}
        \label{fig5}
\end{figure}
Figure \ref{fig5} illustrates the variation of the laser-assisted total cross section of the neutral higgs-boson pair production as a function of the laser field amplitude for different exchanged photons number and for two known laser frequencies which are $\omega=0.117 eV$ ($CO_{2}$ laser) and $\omega=1,17 eV$ (ND:YAG laser). We should mention that the centre of mass energy is chosen as $\sqrt{s}=250\,GeV$ in both cases. According to this figure, for low laser field amplitudes, the total cross section doesn't show any dependence on the laser field amplitude regardless of the number of exchanged photons. However, the threshold value of the laser field strength, from which the laser field begins to affect the cross section, depends on the number of exchanged photons and the laser field frequency. It raises as long as the number of exchanged photons increases or by decreasing the laser frequency. In addition, it is obvious that if the laser field overcomes this threshold value, the total cross section decrease progressively until it becomes zero. Moreover, if we sum over the corresponding cutoff number, we  notice that the laser field will not show its effect anymore. To have a complete picture about how the cross section behaves inside the laser field, we have plotted in one figure its variation as a function of the laser field strength for different centre of mass energies.
\begin{figure}[H]
  \centering
      \includegraphics[scale=0.80]{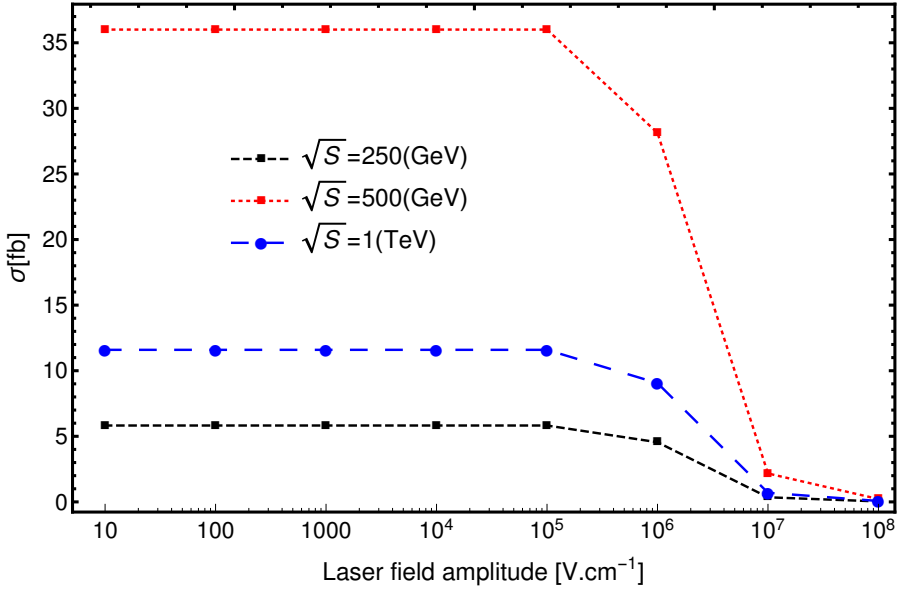}\hspace*{0.4cm}
        \caption{The laser-assisted total cross section of the process ${e}^{+}{e}^{-}\rightarrow H^{0}A^{0}$ versus the laser field strength for different centre of mass energies and by taking the laser frequency and the neutral Higgs-boson mass as: $\omega=1,17 eV$ and $m_{s}=120\,GeV$.}
        \label{fig6}
\end{figure}
Figure \ref{fig6} represents the laser-assisted total cross section as a function of the laser field strength for three typical centre of mass energies which are $250\,GeV$, $500\,GeV$ and $1\,TeV$. The results presented in this figure confirms that obtained in figure (\ref{fig5}b) ($\omega=1,17 eV$) which states that, regardless of the centre of mass energy, the laser field do not affect the total cross section at low laser field strengths ( $\epsilon_{0}\leq 10^{5}V.cm^{-1}$). In addition, the total cross section begins to decrease as long as the laser field strength overcomes $\epsilon_{0}= 10^{5}V.cm^{-1}$. Moreover, though the laser-assisted total cross section has different orders of magnitudes, it has the same general aspects. Specifically, the order of magnitude of the laser-assisted total cross section of the process ${e}^{+}{e}^{-}\rightarrow H^{0}A^{0}$ at the centre of mass energy $\sqrt{s}=500\,GeV$ is approximately six times higher than that obtained in the case where $\sqrt{s}=250\,GeV$, and it is three times higher than that which presents the case of $\sqrt{s}=1\,TeV$. Therefore, we can conclude that the maximum of the cross section is reached for the energies near $\sqrt{s}=500\,GeV$. Another important point to be mentioned here is that, for $m_{s}=120\,GeV$, we remark that, for example, the total cross section reaches $\sigma=36\,[fb]$ which is mach higher than those obtained in figure \ref{fig4}. Therefore, this result confirms that the maximum of the total cross section increases as much as the neutral Higgs-bosons mass decreases.
\section{Conclusion}
In the present paper, we have investigated the neutral Higgs-bosons pair production in the presence of a circularly polarized laser field in the IHDM model without taking into consideration the radiative corrections. We have found that, for a given laser field strength and frequency, the partial total cross section presents two symmetric cutoffs. Then, we have checked the validity of the Kroll-Watson sum-rule, so the summation of the laser-assisted partial total cross section from $-cutoff$ to $+cutoff$ is equal to its corresponding laser-free total cross section. In addition, we have analyzed the dependence of the laser-assisted total cross section on the centre of mass energy, and we have indicated that its order of magnitude decreases by increasing the outgoing neutral Higgs-bosons masses. After that, we have shown that the laser field doesn't affect the total cross section at low laser field strengths. However, whenever the latter overcomes a threshold value which depends on the number of exchanged photons, the total cross section progressively decreases. Furthermore, we have shown that, for a given laser field strength and frequency, the total cross section has a high order of magnitude for $\sqrt{s}=500\,GeV$ as compared to the case where $\sqrt{s}=250\,GeV$ or $\sqrt{s}=1\,TeV$.
\section{Appendix}
\tiny
\begin{eqnarray}
B&=&\nonumber\dfrac{e^2 }{(k.p_{1})(k.p_{2})}\Big[
    (4 (k.p_{1}) (k.p_{2}) ((a_{2}.p_{3})^2 g_{a}^{e^{2}} (k.p_{1}) (k.p_{2}) - 2 (a_{2}.p_{3}) (a_{2}.p_{4}) g_{a}^{e^{2}} (k.p_{1}) (k.p_{2}) + 
        (a_{2}.p_{4})^2 g_{a}^{e^{2}}  (k.p_{1}) (k.p_{2}) + (a_{2}.p_{3})^2 g_{v}^{e^{2}} (k.p_{1}) (k.p_{2}) \\&- &\nonumber 
        2 (a_{2}.p_{3}) (a_{2}.p_{4}) g_{v}^{e^{2}} (k.p_{1}) (k.p_{2}) + (a_{2}.p_{4})^2 g_{v}^{e^{2}} (k.p_{1}) (k.p_{2})   
        (a_{1}.p_{3})^2 (g_{a}^{e^{2}} + g_{v}^{e^{2}}) (k.p_{1}) (k.p_{2}) + 
        (a_{1}.p_{4})^2 (g_{a}^{e^{2}} + g_{v}^{e^{2}}) (k.p_{1}) (k.p_{2}) + (a_{1}.p_{1})\\&\times &\nonumber (a_{1}.p_{2}) g_{a}^{e^{2}} (k.p_{3})^2  
        (a_{1}.p_{1}) (a_{1}.p_{2}) g_{v}^{e^{2}} (k.p_{3})^2 - 
        (a_{1}.p_{3}) (g_{a}^{e^{2}} + 
           g_{v}^{e^{2}}) (2 (a_{1}.p_{4}) (k.p_{1}) (k.p_{2}) + ((a_{1}.p_{2}) (k.p_{1}) + (a_{1}.p_{1})  (k.p_{2})) ((k.p_{3}) - 
              (k.p_{4}))) \\&+ &\nonumber 
        (a_{1}.p_{4}) (g_{a}^{e^{2}} + g_{v}^{e^{2}}) ((a_{1}.p_{2}) (k.p_{1}) + (a_{1}.p_{1}) (k.p_{2})) ((k.p_{3}) - (k.p_{4}))  
        2 (a_{1}.p_{1}) (a_{1}.p_{2}) g_{a}^{e^{2}} (k.p_{3}) (k.p_{4}) - 2 (a_{1}.p_{1}) (a_{1}.p_{2}) g_{v}^{e^{2}} (k.p_{3}) (k.p_{4}) + 
        (a_{1}.p_{1})\\&\times &\nonumber (a_{1}.p_{2}) g_{a}^{e^{2}} (k.p_{4})^2   (a_{1}.p_{1}) (a_{1}.p_{2}) g_{v}^{e^{2}} (k.p_{4})^2 - 
        a^2 g_{a}^{e^{2}} (k.p_{1})^2 m_{A^{0}}^{2} - a^2 g_{v}^{e^{2}} (k.p_{1})^2 m_{A^{0}}^{2} - 
        2 a^2 g_{a}^{e^{2}} (k.p_{1}) (k.p_{2}) m_{A^{0}}^{2}  2 a^2 g_{v}^{e^{2}} (k.p_{1}) (k.p_{2}) m_{A^{0}}^{2} - 
        a^2 \\&\times &\nonumber g_{a}^{e^{2}} (k.p_{2})^2 m_{A^{0}}^{2} - a^2 g_{v}^{e^{2}} (k.p_{2})^2 m_{A^{0}}^{2} + 
        a^2 g_{a}^{e^{2}} (k.p_{3})^2 me^2 - a^2 g_{v}^{e^{2}} (k.p_{3})^2  me^2 - 
        2 a^2 g_{a}^{e^{2}} (k.p_{3}) (k.p_{4}) me^2 + 2 a^2 g_{v}^{e^{2}} (k.p_{3}) (k.p_{4}) me^2 + 
        a^2 g_{a}^{e^{2}}\\&\times &\nonumber (k.p_{4})^2 me^2 - a^2 g_{v}^{e^{2}} (k.p_{4})^2 me^2 
        a^2 g_{a}^{e^{2}} (k.p_{3})^2 (p_{1}.p_{2}) - a^2 g_{v}^{e^{2}} (k.p_{3})^2 (p_{1}.p_{2}) + 
        2 a^2 g_{a}^{e^{2}} (k.p_{3}) (k.p_{4}) (p_{1}.p_{2}) + 2 a^2 g_{v}^{e^{2}} (k.p_{3}) (k.p_{4}) (p_{1}.p_{2})-  
        a^2 g_{a}^{e^{2}} \\&\times &\nonumber (k.p_{4})^2 (p_{1}.p_{2}) - a^2 g_{v}^{e^{2}} (k.p_{4})^2 (p_{1}.p_{2}) + 
        a^2 g_{a}^{e^{2}} (k.p_{1}) (k.p_{3}) (p_{1}.p_{3}) + a^2 g_{v}^{e^{2}} (k.p_{1}) (k.p_{3}) (p_{1}.p_{3}) +  
        a^2 g_{a}^{e^{2}} (k.p_{2}) (k.p_{3}) (p_{1}.p_{3}) + a^2 g_{v}^{e^{2}} (k.p_{2}) (k.p_{3})\\&\times &\nonumber (p_{1}.p_{3}) - 
        a^2 g_{a}^{e^{2}} (k.p_{1}) (k.p_{4}) (p_{1}.p_{3}) - a^2 g_{v}^{e^{2}} (k.p_{1})  (k.p_{4}) (p_{1}.p_{3}) - 
        a^2 g_{a}^{e^{2}} (k.p_{2}) (k.p_{4}) (p_{1}.p_{3}) - a^2 g_{v}^{e^{2}} (k.p_{2}) (k.p_{4}) (p_{1}.p_{3}) - 
        a^2 g_{a}^{e^{2}} (k.p_{1}) (k.p_{3})\\&\times &\nonumber (p_{1}.p_{4}) - a^2 g_{v}^{e^{2}} (k.p_{1}) (k.p_{3}) (p_{1}.p_{4}) - 
        a^2 g_{a}^{e^{2}} (k.p_{2}) (k.p_{3}) (p_{1}.p_{4}) - a^2 g_{v}^{e^{2}} (k.p_{2}) (k.p_{3}) (p_{1}.p_{4}) + 
        a^2 g_{a}^{e^{2}} (k.p_{1})  (k.p_{4}) (p_{1}.p_{4}) + a^2 g_{v}^{e^{2}} (k.p_{1}) (k.p_{4})\\&\times &\nonumber (p_{1}.p_{4}) + 
        a^2 g_{a}^{e^{2}} (k.p_{2}) (k.p_{4})(p_{1}.p_{4}) + a^2 g_{v}^{e^{2}} (k.p_{2}) (k.p_{4}) (p_{1}.p_{4}) +  
        a^2 g_{a}^{e^{2}} (k.p_{1}) (k.p_{3}) (p_{2}.p_{3}) + a^2 g_{v}^{e^{2}} (k.p_{1}) (k.p_{3}) (p_{2}.p_{3}) + 
        a^2 g_{a}^{e^{2}} (k.p_{2}) (k.p_{3})\\&\times &\nonumber (p_{2}.p_{3}) + a^2 g_{v}^{e^{2}} (k.p_{2}) (k.p_{3}) (p_{2}.p_{3}) - 
        a^2 g_{a}^{e^{2}} (k.p_{1}) (k.p_{4}) (p_{2}.p_{3}) - a^2 g_{v}^{e^{2}} (k.p_{1}) (k.p_{4}) (p_{2}.p_{3}) - 
        a^2 g_{a}^{e^{2}} (k.p_{2}) (k.p_{4}) (p_{2}.p_{3}) - a^2 g_{v}^{e^{2}}  (k.p_{2}) (k.p_{4})\\&\times &\nonumber (p_{2}.p_{3}) - 
        a^2 g_{a}^{e^{2}} (k.p_{1}) (k.p_{3}) (p_{2}.p_{4}) - a^2 g_{v}^{e^{2}} (k.p_{1}) (k.p_{3}) (p_{2}.p_{4}) - 
        a^2 g_{a}^{e^{2}} (k.p_{2}) (k.p_{3})  (p_{2}.p_{4}) - a^2 g_{v}^{e^{2}} (k.p_{2}) (k.p_{3}) (p_{2}.p_{4}) + 
        a^2 g_{a}^{e^{2}} (k.p_{1}) (k.p_{4})\\&\times &\nonumber (p_{2}.p_{4}) + a^2 g_{v}^{e^{2}} (k.p_{1}) (k.p_{4}) (p_{2}.p_{4}) + 
        a^2 g_{a}^{e^{2}}  (k.p_{2}) (k.p_{4}) (p_{2}.p_{4}) + a^2 g_{v}^{e^{2}} (k.p_{2}) (k.p_{4}) (p_{2}.p_{4}) + 
        a^2 (g_{a}^{e^{2}} + g_{v}^{e^{2}}) ((k.p_{1}) + (k.p_{2}))^2 (p_{3}.p_{4})) + 
     g_{a}^{e} g_{v}^{e}\\&\times &\nonumber (k.p_{2}) ((k.p_{2}) (-3 ((k.p_{3}) - (k.p_{4})) ((p_{2}.p_{3}) - (p_{2}.p_{4})) + 
           4 (k.p_{2}) (m_{A^{0}}^{2} - (p_{3}.p_{4}))) + 
        (k.p_{1}) (-((k.p_{3}) - (k.p_{4})) ((p_{2}.p_{3}) - (p_{2}.p_{4})) + 8 (k.p_{2})\\&\times &\nonumber (m_{A^{0}}^{2} - (p_{3}.p_{4}))) + 
        4 (k.p_{1})^2 (m_{A^{0}}^{2} - (p_{3}.p_{4}))) \epsilon(a_{1}, a_{2}, 
       k, p_{1}) + 
     g_{a}^{e} g_{v}^{e} (k.p_{1})  (4 (k.p_{1})^2 (-m_{A^{0}}^{2} + (p_{3}.p_{4})) + 
        (k.p_{2}) (((k.p_{3}) - (k.p_{4})) ((p_{1}.p_{3}) \\&- &\nonumber(p_{1}.p_{4})) + 4 (k.p_{2}) (-m_{A^{0}}^{2} + (p_{3}.p_{4})))  
        (k.p_{1}) (3 ((k.p_{3}) - (k.p_{4})) ((p_{1}.p_{3}) - (p_{1}.p_{4})) + 8 (k.p_{2}) (-m_{A^{0}}^{2} + (p_{3}.p_{4})))) \epsilon(
       a_{1}, a_{2}, k, p_{2}) + 
     g_{a}^{e} g_{v}^{e} (((k.p_{1}) \\&- &\nonumber (k.p_{2})) ((k.p_{1}) + (k.p_{2})) ((k.p_{3}) - (k.p_{4})) (p_{1}.p_{2}) \epsilon(
          a_{1}, a_{2}, k, 
          p_{3}) - ((k.p_{1}) - (k.p_{2})) ((k.p_{1}) + (k.p_{2})) ((k.p_{3}) - 
           (k.p_{4})) (p_{1}.p_{2}) \epsilon(a_{1}, a_{2}, k, 
          p_{4}) \\&+ &\nonumber 
        3 (k.p_{1})^2 (k.p_{3})^2 \epsilon(a_{1}, a_{2}, p_{1}, 
          p_{2}) - 
        6 (k.p_{1}) (k.p_{2})  (k.p_{3})^2 \epsilon(a_{1}, a_{2}, p_{1}, 
          p_{2}) + 
        3 (k.p_{2})^2 (k.p_{3})^2 \epsilon(a_{1}, a_{2}, p_{1}, 
          p_{2}) - 
        6 (k.p_{1})^2 (k.p_{3}) (k.p_{4}) \epsilon(a_{1}, a_{2}, p_{1}, 
          p_{2}) \\&+ &\nonumber  
        12 (k.p_{1}) (k.p_{2}) (k.p_{3}) (k.p_{4}) \epsilon(a_{1}, a_{2}, 
          p_{1}, p_{2}) - 
        6 (k.p_{2})^2 (k.p_{3}) (k.p_{4}) \epsilon(a_{1}, a_{2}, p_{1}, 
          p_{2}) + 
        3 (k.p_{1})^2 (k.p_{4})^2  \epsilon(a_{1}, a_{2}, p_{1}, 
          p_{2}) - 
        6 (k.p_{1}) (k.p_{2}) (k.p_{4})^2\\&\times &\nonumber \epsilon(a_{1}, a_{2}, p_{1}, 
          p_{2}) + 
        3 (k.p_{2})^2 (k.p_{4})^2 \epsilon(a_{1}, a_{2}, p_{1}, 
          p_{2}) + 
        (k.p_{1})^2 (k.p_{2}) (k.p_{3}) \epsilon(a_{1}, a_{2}, p_{1}, 
          p_{3}) + 
        5 (k.p_{1}) (k.p_{2})^2 (k.p_{3}) \epsilon(a_{1}, a_{2}, p_{1}, 
          p_{3}) - 
        2 (k.p_{2})^3 (k.p_{3})\\&\times &\nonumber \epsilon(a_{1}, a_{2}, p_{1}, 
          p_{3}) - 
        (k.p_{1})^2 (k.p_{2}) (k.p_{4}) \epsilon(a_{1}, a_{2}, p_{1}, 
          p_{3}) - 
        5 (k.p_{1}) (k.p_{2})^2 (k.p_{4}) \epsilon(a_{1}, a_{2}, p_{1}, 
          p_{3}) + 
        2 (k.p_{2})^3 (k.p_{4})  \epsilon(a_{1}, a_{2}, p_{1}, 
          p_{3}) - 
        (k.p_{1})^2 (k.p_{2})\\&\times &\nonumber (k.p_{3}) \epsilon(a_{1}, a_{2}, p_{1}, 
          p_{4}) - 
        5 (k.p_{1}) (k.p_{2})^2 (k.p_{3}) \epsilon(a_{1}, a_{2}, p_{1}, 
          p_{4}) +
        2 (k.p_{2})^3 (k.p_{3}) \epsilon(a_{1}, a_{2}, p_{1}, 
          p_{4}) + 
        (k.p_{1})^2 (k.p_{2}) (k.p_{4}) \epsilon(a_{1}, a_{2}, p_{1}, 
          p_{4}) + 
        5 (k.p_{1})\\&\times &\nonumber (k.p_{2})^2 (k.p_{4})  \epsilon(a_{1}, a_{2}, p_{1}, 
          p_{4}) - 
        2 (k.p_{2})^3 (k.p_{4}) \epsilon(a_{1}, a_{2}, p_{1}, 
          p_{4}) + 
        2 (k.p_{1})^3 (k.p_{3}) \epsilon(a_{1}, a_{2}, p_{2}, 
          p_{3}) - 
        5 (k.p_{1})^2 (k.p_{2})  (k.p_{3}) \epsilon(a_{1}, a_{2}, p_{2}, 
          p_{3}) - 
        (k.p_{1}) \\&\times &\nonumber(k.p_{2})^2 (k.p_{3}) \epsilon(a_{1}, a_{2}, p_{2}, 
          p_{3}) - 
        2 (k.p_{1})^3 (k.p_{4}) \epsilon(a_{1}, a_{2}, p_{2}, 
          p_{3}) + 
        5 (k.p_{1})^2  (k.p_{2}) (k.p_{4}) \epsilon(a_{1}, a_{2}, p_{2}, 
          p_{3}) + 
        (k.p_{1}) (k.p_{2})^2 (k.p_{4}) \epsilon(a_{1}, a_{2}, p_{2}, 
          p_{3}) \\&- &\nonumber
        2 (k.p_{1})^3 (k.p_{3}) \epsilon(a_{1}, a_{2}, p_{2}, 
          p_{4}) + 
        5 (k.p_{1})^2 (k.p_{2}) (k.p_{3}) \epsilon(a_{1}, a_{2}, p_{2}, 
          p_{4}) + 
        (k.p_{1}) (k.p_{2})^2 (k.p_{3}) \epsilon(a_{1}, a_{2}, p_{2}, 
          p_{4}) + 
        2 (k.p_{1})^3 (k.p_{4}) \epsilon(a_{1}, a_{2}, p_{2}, 
          p_{4}) \\&- &\nonumber
        5 (k.p_{1})^2 (k.p_{2}) (k.p_{4}) \epsilon(a_{1}, a_{2}, p_{2}, 
          p_{4}) - 
        (k.p_{1}) (k.p_{2})^2 (k.p_{4}) \epsilon(a_{1}, a_{2}, p_{2}, 
          p_{4}) + 
        4 (a_{2}.p_{3}) (k.p_{1}) (k.p_{2}) (k.p_{3}) \epsilon(a_{1}, k, 
          p_{1}, p_{2}) - 
        4 (a_{2}.p_{4}) (k.p_{1}) (k.p_{2})\\&\times &\nonumber (k.p_{3}) \epsilon(a_{1}, k, 
          p_{1}, p_{2}) - 
        4 (a_{2}.p_{3}) (k.p_{1}) (k.p_{2}) (k.p_{4}) \epsilon(a_{1}, k, 
          p_{1}, p_{2}) + 
        4 (a_{2}.p_{4}) (k.p_{1}) (k.p_{2}) (k.p_{4}) \epsilon(a_{1}, k, 
          p_{1}, p_{2}) - 
        4 (a_{2}.p_{3}) (k.p_{1})^2 (k.p_{2})\\&\times &\nonumber  \epsilon(a_{1}, k, p_{1}, 
          p_{3}) + 
        4 (a_{2}.p_{4}) (k.p_{1})^2 (k.p_{2}) \epsilon(a_{1}, k, p_{1}, 
          p_{3}) - 
        4 (a_{2}.p_{3}) (k.p_{1}) (k.p_{2})^2 \epsilon(a_{1}, k, p_{1}, 
          p_{3}) +  
        4 (a_{2}.p_{4}) (k.p_{1}) (k.p_{2})^2 \epsilon(a_{1}, k, p_{1}, 
          p_{3}) + 
        4 (a_{2}.p_{3})\\&\times &\nonumber (k.p_{1})^2 (k.p_{2}) \epsilon(a_{1}, k, p_{1}, 
          p_{4}) - 
        4 (a_{2}.p_{4}) (k.p_{1})^2 (k.p_{2})  \epsilon(a_{1}, k, p_{1}, 
          p_{4}) + 
        4 (a_{2}.p_{3}) (k.p_{1}) (k.p_{2})^2 \epsilon(a_{1}, k, p_{1}, 
          p_{4}) - 
        4 (a_{2}.p_{4}) (k.p_{1}) (k.p_{2})^2 \epsilon(a_{1}, k, p_{1}, 
          p_{4}) \\&+ &\nonumber 
        4 (a_{2}.p_{3}) (k.p_{1})^2 (k.p_{2}) \epsilon(a_{1}, k, p_{2}, 
          p_{3}) - 
        4 (a_{2}.p_{4}) (k.p_{1})^2 (k.p_{2}) \epsilon(a_{1}, k, p_{2}, 
          p_{3}) + 
        4 (a_{2}.p_{3}) (k.p_{1}) (k.p_{2})^2  \epsilon(a_{1}, k, p_{2}, 
          p_{3}) - 
        4 (a_{2}.p_{4}) (k.p_{1}) (k.p_{2})^2\\&\times &\nonumber \epsilon(a_{1}, k, p_{2}, 
          p_{3}) - 
        4 (a_{2}.p_{3}) (k.p_{1})^2 (k.p_{2}) \epsilon(a_{1}, k, p_{2}, 
          p_{4}) + 
        4 (a_{2}.p_{4}) (k.p_{1})^2 (k.p_{2}) \epsilon(a_{1}, k, p_{2}, 
          p_{4}) - 
        4 (a_{2}.p_{3}) (k.p_{1}) (k.p_{2})^2 \epsilon(a_{1}, k, p_{2}, 
          p_{4}) + 
        4 (a_{2}.p_{4})\\&\times &\nonumber (k.p_{1}) (k.p_{2})^2 \epsilon(a_{1}, k, p_{2}, 
          p_{4}) -  
        4 (a_{1}.p_{3}) (k.p_{1}) (k.p_{2}) (k.p_{3}) \epsilon(a_{2}, k, 
          p_{1}, p_{2}) + 
        4 (a_{1}.p_{4}) (k.p_{1}) (k.p_{2}) (k.p_{3}) \epsilon(a_{2}, k, 
          p_{1}, p_{2}) + 
        4 (a_{1}.p_{3}) (k.p_{1}) (k.p_{2})\\&\times &\nonumber  (k.p_{4}) \epsilon(a_{2}, k, 
          p_{1}, p_{2}) - 
        4 (a_{1}.p_{4}) (k.p_{1}) (k.p_{2}) (k.p_{4}) \epsilon(a_{2}, k, 
          p_{1}, p_{2}) + 
        4 (a_{1}.p_{3}) (k.p_{1})^2 (k.p_{2}) \epsilon(a_{2}, k, p_{1}, 
          p_{3}) - 
        4 (a_{1}.p_{4}) (k.p_{1})^2 (k.p_{2}) \epsilon(a_{2}, k, p_{1}, 
          p_{3}) \\&+ &\nonumber 
        4 (a_{1}.p_{3}) (k.p_{1}) (k.p_{2})^2 \epsilon(a_{2}, k, p_{1}, 
          p_{3}) - 
        4 (a_{1}.p_{4}) (k.p_{1}) (k.p_{2})^2  \epsilon(a_{2}, k, p_{1}, 
          p_{3}) - 
        (a_{1}.p_{2}) (k.p_{1}) (k.p_{2}) (k.p_{3}) \epsilon(a_{2}, k, p_{1}, 
          p_{3}) + 
        (a_{1}.p_{2}) (k.p_{2})^2 (k.p_{3})\\&\times &\nonumber \epsilon(a_{2}, k, p_{1}, 
          p_{3}) + 
        (a_{1}.p_{2})  (k.p_{1}) (k.p_{2}) (k.p_{4}) \epsilon(a_{2}, k, p_{1}, 
          p_{3}) - 
        (a_{1}.p_{2}) (k.p_{2})^2 (k.p_{4}) \epsilon(a_{2}, k, p_{1}, 
          p_{3}) - 
        4 (a_{1}.p_{3}) (k.p_{1})^2 (k.p_{2}) \epsilon(a_{2}, k, p_{1}, 
          p_{4}) +  
        4 (a_{1}.p_{4})\\&\times &\nonumber (k.p_{1})^2 (k.p_{2}) \epsilon(a_{2}, k, p_{1}, 
          p_{4}) - 
        4 (a_{1}.p_{3}) (k.p_{1}) (k.p_{2})^2 \epsilon(a_{2}, k, p_{1}, 
          p_{4}) + 
        4 (a_{1}.p_{4}) (k.p_{1}) (k.p_{2})^2 \epsilon(a_{2}, k, p_{1}, 
          p_{4}) +  
        (a_{1}.p_{2}) (k.p_{1}) (k.p_{2}) (k.p_{3})\\&\times &\nonumber \epsilon(a_{2}, k, p_{1}, 
          p_{4}) - 
        (a_{1}.p_{2}) (k.p_{2})^2 (k.p_{3}) \epsilon(a_{2}, k, p_{1}, 
          p_{4}) - 
        (a_{1}.p_{2}) (k.p_{1}) (k.p_{2}) (k.p_{4}) \epsilon(a_{2}, k, p_{1}, 
          p_{4}) +  
        (a_{1}.p_{2}) (k.p_{2})^2 (k.p_{4}) \epsilon(a_{2}, k, p_{1}, 
          p_{4}) - 
        4 (a_{1}.p_{3})\\&\times &\nonumber (k.p_{1})^2 (k.p_{2}) \epsilon(a_{2}, k, p_{2}, 
          p_{3}) + 
        4 (a_{1}.p_{4}) (k.p_{1})^2 (k.p_{2}) \epsilon(a_{2}, k, p_{2}, 
          p_{3}) - 
        4 (a_{1}.p_{3}) (k.p_{1}) (k.p_{2})^2 \epsilon(a_{2}, k, p_{2}, 
          p_{3}) + 
        4 (a_{1}.p_{4}) (k.p_{1}) (k.p_{2})^2 \epsilon(a_{2}, k, p_{2}, 
          p_{3}) \\&- &\nonumber 
        (a_{1}.p_{1}) (k.p_{1})^2 (k.p_{3}) \epsilon(a_{2}, k, p_{2}, 
          p_{3}) + 
        (a_{1}.p_{1}) (k.p_{1}) (k.p_{2}) (k.p_{3}) \epsilon(a_{2}, k, p_{2}, 
          p_{3}) + 
        (a_{1}.p_{1}) (k.p_{1})^2 (k.p_{4}) \epsilon(a_{2}, k, p_{2}, 
          p_{3}) - 
        (a_{1}.p_{1}) (k.p_{1}) (k.p_{2}) (k.p_{4})\\&\times &\nonumber \epsilon(a_{2}, k, p_{2}, 
          p_{3}) + 
        (k.p_{1})  (4 (a_{1}.p_{3}) (k.p_{2}) ((k.p_{1}) + (k.p_{2})) - 4 (a_{1}.p_{4}) (k.p_{2}) ((k.p_{1}) + (k.p_{2})) + 
           (a_{1}.p_{1}) ((k.p_{1}) - (k.p_{2})) ((k.p_{3}) - (k.p_{4}))) \epsilon(a_{2}, 
          k, p_{2}, p_{4})))\Big]
\end{eqnarray}
\normalsize
\tiny
\begin{eqnarray}
C&=&\nonumber\dfrac{-e^2}{(2 (k.p_{1})^2 (k.p_{2})^2)}
    \Big[(-4 (k.p_{1}) (k.p_{2}) (-(a_{2}.p_{3})^2 g_{a}^{e^{2}} (k.p_{1}) (k.p_{2}) + 
        2 (a_{2}.p_{3}) (a_{2}.p_{4}) g_{a}^{e^{2}} (k.p_{1}) (k.p_{2}) - (a_{2}.p_{4})^2 g_{a}^{e^{2}} (k.p_{1}) (k.p_{2}) - 
        (a_{2}.p_{3})^2 g_{v}^{e^{2}} (k.p_{1})\\&\times &\nonumber (k.p_{2}) + 2 (a_{2}.p_{3}) (a_{2}.p_{4}) g_{v}^{e^{2}} (k.p_{1}) (k.p_{2}) - 
        (a_{2}.p_{4})^2 g_{v}^{e^{2}} (k.p_{1}) (k.p_{2}) - (a_{1}.p_{3})^2 (g_{a}^{e^{2}} + g_{v}^{e^{2}}) (k.p_{1}) (k.p_{2}) - 
        (a_{1}.p_{4})^2 (g_{a}^{e^{2}} + g_{v}^{e^{2}}) (k.p_{1}) (k.p_{2}) \\&- &\nonumber (a_{1}.p_{1}) (a_{1}.p_{2}) g_{a}^{e^{2}} (k.p_{3})^2 - 
        (a_{1}.p_{1}) (a_{1}.p_{2}) g_{v}^{e^{2}} (k.p_{3})^2 + 
        (a_{1}.p_{3}) (g_{a}^{e^{2}} + 
           g_{v}^{e^{2}}) (2 (a_{1}.p_{4}) (k.p_{1}) (k.p_{2}) + ((a_{1}.p_{2}) (k.p_{1}) + (a_{1}.p_{1}) (k.p_{2})) ((k.p_{3}) \\&- &\nonumber 
              (k.p_{4}))) - 
        (a_{1}.p_{4}) (g_{a}^{e^{2}} + g_{v}^{e^{2}}) ((a_{1}.p_{2}) (k.p_{1}) + (a_{1}.p_{1}) (k.p_{2})) ((k.p_{3}) - (k.p_{4})) + 
        2 (a_{1}.p_{1}) (a_{1}.p_{2}) g_{a}^{e^{2}} (k.p_{3}) (k.p_{4}) + 2 (a_{1}.p_{1}) (a_{1}.p_{2}) g_{v}^{e^{2}} (k.p_{3})\\&\times &\nonumber (k.p_{4}) - 
        (a_{1}.p_{1}) (a_{1}.p_{2}) g_{a}^{e^{2}} (k.p_{4})^2 - (a_{1}.p_{1}) (a_{1}.p_{2}) g_{v}^{e^{2}} (k.p_{4})^2 + 
        a^2 g_{a}^{e^{2}} (k.p_{1})^2 m_{A^{0}}^{2} + a^2 g_{v}^{e^{2}} (k.p_{1})^2 m_{A^{0}}^{2} + 
        2 a^2 g_{a}^{e^{2}} (k.p_{1}) (k.p_{2}) m_{A^{0}}^{2} + 2 a^2 g_{v}^{e^{2}} \\&\times &\nonumber(k.p_{1}) (k.p_{2}) m_{A^{0}}^{2} + 
        a^2 g_{a}^{e^{2}} (k.p_{2})^2 m_{A^{0}}^{2} + a^2 g_{v}^{e^{2}} (k.p_{2})^2 m_{A^{0}}^{2} - 
        a^2 g_{a}^{e^{2}} (k.p_{3})^2 me^2 + a^2 g_{v}^{e^{2}} (k.p_{3})^2 me^2 + 
        2 a^2 g_{a}^{e^{2}} (k.p_{3}) (k.p_{4}) me^2 - 2 a^2 g_{v}^{e^{2}} (k.p_{3})\\&\times &\nonumber (k.p_{4}) me^2 - 
        a^2 g_{a}^{e^{2}} (k.p_{4})^2 me^2 + a^2 g_{v}^{e^{2}} (k.p_{4})^2 me^2 + 
        a^2 g_{a}^{e^{2}} (k.p_{3})^2 (p_{1}.p_{2}) + a^2 g_{v}^{e^{2}} (k.p_{3})^2 (p_{1}.p_{2}) - 
        2 a^2 g_{a}^{e^{2}} (k.p_{3}) (k.p_{4}) (p_{1}.p_{2}) - 2 a^2 g_{v}^{e^{2}} (k.p_{3})\\&\times &\nonumber (k.p_{4}) (p_{1}.p_{2}) + 
        a^2 g_{a}^{e^{2}} (k.p_{4})^2 (p_{1}.p_{2}) + a^2 g_{v}^{e^{2}} (k.p_{4})^2 (p_{1}.p_{2}) - 
        a^2 g_{a}^{e^{2}} (k.p_{1}) (k.p_{3}) (p_{1}.p_{3}) - a^2 g_{v}^{e^{2}} (k.p_{1}) (k.p_{3}) (p_{1}.p_{3}) - 
        a^2 g_{a}^{e^{2}} (k.p_{2}) (k.p_{3}) (p_{1}.p_{3})\\&-&\nonumber a^2 g_{v}^{e^{2}} (k.p_{2}) (k.p_{3}) (p_{1}.p_{3}) + 
        a^2 g_{a}^{e^{2}} (k.p_{1}) (k.p_{4}) (p_{1}.p_{3}) + a^2 g_{v}^{e^{2}} (k.p_{1}) (k.p_{4}) (p_{1}.p_{3}) + 
        a^2 g_{a}^{e^{2}} (k.p_{2}) (k.p_{4}) (p_{1}.p_{3}) + a^2 g_{v}^{e^{2}} (k.p_{2}) (k.p_{4}) (p_{1}.p_{3}) + 
        a^2 g_{a}^{e^{2}}\\&\times &\nonumber (k.p_{1}) (k.p_{3}) (p_{1}.p_{4}) + a^2 g_{v}^{e^{2}} (k.p_{1}) (k.p_{3}) (p_{1}.p_{4}) + 
        a^2 g_{a}^{e^{2}} (k.p_{2}) (k.p_{3}) (p_{1}.p_{4}) + a^2 g_{v}^{e^{2}} (k.p_{2}) (k.p_{3}) (p_{1}.p_{4}) - 
        a^2 g_{a}^{e^{2}} (k.p_{1}) (k.p_{4}) (p_{1}.p_{4}) - a^2 g_{v}^{e^{2}} \\&\times &\nonumber(k.p_{1}) (k.p_{4}) (p_{1}.p_{4}) - 
        a^2 g_{a}^{e^{2}} (k.p_{2}) (k.p_{4}) (p_{1}.p_{4}) - a^2 g_{v}^{e^{2}} (k.p_{2}) (k.p_{4}) (p_{1}.p_{4}) - 
        a^2 g_{a}^{e^{2}} (k.p_{1}) (k.p_{3}) (p_{2}.p_{3}) - a^2 g_{v}^{e^{2}} (k.p_{1}) (k.p_{3}) (p_{2}.p_{3}) - 
        a^2 g_{a}^{e^{2}} \\&\times &\nonumber(k.p_{2}) (k.p_{3}) (p_{2}.p_{3}) - a^2 g_{v}^{e^{2}} (k.p_{2}) (k.p_{3}) (p_{2}.p_{3}) + 
        a^2 g_{a}^{e^{2}} (k.p_{1}) (k.p_{4}) (p_{2}.p_{3}) + a^2 g_{v}^{e^{2}} (k.p_{1}) (k.p_{4}) (p_{2}.p_{3}) + 
        a^2 g_{a}^{e^{2}} (k.p_{2}) (k.p_{4}) (p_{2}.p_{3}) + a^2 g_{v}^{e^{2}} \\&\times &\nonumber(k.p_{2}) (k.p_{4}) (p_{2}.p_{3}) + 
        a^2 g_{a}^{e^{2}} (k.p_{1}) (k.p_{3}) (p_{2}.p_{4}) + a^2 g_{v}^{e^{2}} (k.p_{1}) (k.p_{3}) (p_{2}.p_{4}) + 
        a^2 g_{a}^{e^{2}} (k.p_{2}) (k.p_{3}) (p_{2}.p_{4}) + a^2 g_{v}^{e^{2}} (k.p_{2}) (k.p_{3}) (p_{2}.p_{4}) - 
        a^2 g_{a}^{e^{2}} \\&\times &\nonumber(k.p_{1}) (k.p_{4}) (p_{2}.p_{4}) - a^2 g_{v}^{e^{2}} (k.p_{1}) (k.p_{4}) (p_{2}.p_{4}) - 
        a^2 g_{a}^{e^{2}} (k.p_{2}) (k.p_{4}) (p_{2}.p_{4}) - a^2 g_{v}^{e^{2}} (k.p_{2}) (k.p_{4}) (p_{2}.p_{4}) - 
        a^2 (g_{a}^{e^{2}} + g_{v}^{e^{2}}) ((k.p_{1}) + (k.p_{2}))^2 (p_{3}.p_{4})) \\&+ &\nonumber 
     g_{a}^{e} g_{v}^{e} (k.p_{2}) (4 (k.p_{1})^2 (-m_{A^{0}}^{2} + (p_{3}.p_{4})) + 
        (k.p_{2}) (3 ((k.p_{3}) - (k.p_{4})) ((p_{2}.p_{3}) - (p_{2}.p_{4})) + 4 (k.p_{2}) (-m_{A^{0}}^{2} + (p_{3}.p_{4}))) + 
        (k.p_{1}) (((k.p_{3}) - (k.p_{4}))\\&\times &\nonumber ((p_{2}.p_{3}) - (p_{2}.p_{4})) + 8 (k.p_{2}) (-m_{A^{0}}^{2} + (p_{3}.p_{4})))) \epsilon(
       a_{1}, a_{2}, k, p_{1}) + 
     g_{a}^{e} g_{v}^{e} (k.p_{1}) ((k.p_{2}) (-((k.p_{3}) - (k.p_{4})) ((p_{1}.p_{3}) - (p_{1}.p_{4})) + 
           4 (k.p_{2}) (m_{A^{0}}^{2} \\&- &\nonumber (p_{3}.p_{4}))) + 
        (k.p_{1}) (-3 ((k.p_{3}) - (k.p_{4})) ((p_{1}.p_{3}) - (p_{1}.p_{4})) + 8 (k.p_{2}) (m_{A^{0}}^{2} - (p_{3}.p_{4}))) + 
        4 (k.p_{1})^2 (m_{A^{0}}^{2} - (p_{3}.p_{4}))) \epsilon(a_{1}, a_{2}, 
       k, p_{2}) + 
     g_{a}^{e} g_{v}^{e}\\&\times &\nonumber (-((k.p_{1}) - (k.p_{2})) ((k.p_{1}) + (k.p_{2})) ((k.p_{3}) - (k.p_{4})) (p_{1}.p_{2}) \epsilon(
          a_{1}, a_{2}, k, 
          p_{3}) + ((k.p_{1}) - (k.p_{2})) ((k.p_{1}) + (k.p_{2})) ((k.p_{3}) - 
           (k.p_{4})) (p_{1}.p_{2}) \\&\times &\nonumber\epsilon(a_{1}, a_{2}, k, 
          p_{4}) - 
        3 (k.p_{1})^2 (k.p_{3})^2 \epsilon(a_{1}, a_{2}, p_{1}, 
          p_{2}) + 
        6 (k.p_{1}) (k.p_{2}) (k.p_{3})^2 \epsilon(a_{1}, a_{2}, p_{1}, 
          p_{2}) - 
        3 (k.p_{2})^2 (k.p_{3})^2 \epsilon(a_{1}, a_{2}, p_{1}, 
          p_{2}) + 
        6 (k.p_{1})^2 (k.p_{3}) (k.p_{4})\\&\times &\nonumber \epsilon(a_{1}, a_{2}, p_{1}, 
          p_{2}) - 
        12 (k.p_{1}) (k.p_{2}) (k.p_{3}) (k.p_{4}) \epsilon(a_{1}, a_{2}, 
          p_{1}, p_{2}) + 
        6 (k.p_{2})^2 (k.p_{3}) (k.p_{4}) \epsilon(a_{1}, a_{2}, p_{1}, 
          p_{2}) - 
        3 (k.p_{1})^2 (k.p_{4})^2 \epsilon(a_{1}, a_{2}, p_{1}, 
          p_{2}) + 
        6 (k.p_{1}) \\&\times &\nonumber(k.p_{2}) (k.p_{4})^2 \epsilon(a_{1}, a_{2}, p_{1}, 
          p_{2}) - 
        3 (k.p_{2})^2 (k.p_{4})^2 \epsilon(a_{1}, a_{2}, p_{1}, 
          p_{2}) - 
        (k.p_{1})^2 (k.p_{2}) (k.p_{3}) \epsilon(a_{1}, a_{2}, p_{1}, 
          p_{3}) - 
        5 (k.p_{1}) (k.p_{2})^2 (k.p_{3}) \epsilon(a_{1}, a_{2}, p_{1}, 
          p_{3}) \\&+ &\nonumber
        2 (k.p_{2})^3 (k.p_{3}) \epsilon(a_{1}, a_{2}, p_{1}, 
          p_{3}) + 
        (k.p_{1})^2 (k.p_{2}) (k.p_{4}) \epsilon(a_{1}, a_{2}, p_{1}, 
          p_{3}) + 
        5 (k.p_{1}) (k.p_{2})^2 (k.p_{4}) \epsilon(a_{1}, a_{2}, p_{1}, 
          p_{3}) - 
        2 (k.p_{2})^3 (k.p_{4}) \epsilon(a_{1}, a_{2}, p_{1}, 
          p_{3}) \\&+ &\nonumber 
        (k.p_{1})^2 (k.p_{2}) (k.p_{3}) \epsilon(a_{1}, a_{2}, p_{1}, 
          p_{4}) + 
        5 (k.p_{1}) (k.p_{2})^2 (k.p_{3}) \epsilon(a_{1}, a_{2}, p_{1}, 
          p_{4}) - 
        2 (k.p_{2})^3 (k.p_{3}) \epsilon(a_{1}, a_{2}, p_{1}, 
          p_{4}) - 
        (k.p_{1})^2 (k.p_{2}) (k.p_{4}) \epsilon(a_{1}, a_{2}, p_{1}, 
          p_{4}) \\&- &\nonumber 
        5 (k.p_{1}) (k.p_{2})^2 (k.p_{4}) \epsilon(a_{1}, a_{2}, p_{1}, 
          p_{4}) + 
        2 (k.p_{2})^3 (k.p_{4}) \epsilon(a_{1}, a_{2}, p_{1}, 
          p_{4}) - 
        2 (k.p_{1})^3 (k.p_{3}) \epsilon(a_{1}, a_{2}, p_{2}, 
          p_{3}) + 
        5 (k.p_{1})^2 (k.p_{2}) (k.p_{3}) \epsilon(a_{1}, a_{2}, p_{2}, 
          p_{3}) \\&+ &\nonumber 
        (k.p_{1}) (k.p_{2})^2 (k.p_{3}) \epsilon(a_{1}, a_{2}, p_{2}, 
          p_{3}) + 
        2 (k.p_{1})^3 (k.p_{4}) \epsilon(a_{1}, a_{2}, p_{2}, 
          p_{3}) - 
        5 (k.p_{1})^2 (k.p_{2}) (k.p_{4}) \epsilon(a_{1}, a_{2}, p_{2}, 
          p_{3}) - 
        (k.p_{1}) (k.p_{2})^2 (k.p_{4}) \epsilon(a_{1}, a_{2}, p_{2}, 
          p_{3}) \\&+ &\nonumber 
        2 (k.p_{1})^3 (k.p_{3}) \epsilon(a_{1}, a_{2}, p_{2}, 
          p_{4}) - 
        5 (k.p_{1})^2 (k.p_{2}) (k.p_{3}) \epsilon(a_{1}, a_{2}, p_{2}, 
          p_{4}) - 
        (k.p_{1}) (k.p_{2})^2 (k.p_{3}) \epsilon(a_{1}, a_{2}, p_{2}, 
          p_{4}) - 
        2 (k.p_{1})^3 (k.p_{4}) \epsilon(a_{1}, a_{2}, p_{2}, 
          p_{4})\\&+ &\nonumber 
        5 (k.p_{1})^2 (k.p_{2}) (k.p_{4}) \epsilon(a_{1}, a_{2}, p_{2}, 
          p_{4}) + 
        (k.p_{1}) (k.p_{2})^2 (k.p_{4}) \epsilon(a_{1}, a_{2}, p_{2}, 
          p_{4}) - 
        4 (a_{2}.p_{3}) (k.p_{1}) (k.p_{2}) (k.p_{3}) \epsilon(a_{1}, k, 
          p_{1}, p_{2}) + 
        4 (a_{2}.p_{4}) (k.p_{1}) (k.p_{2}) (k.p_{3})\\&\times &\nonumber \epsilon(a_{1}, k, 
          p_{1}, p_{2}) + 
        4 (a_{2}.p_{3}) (k.p_{1}) (k.p_{2}) (k.p_{4}) \epsilon(a_{1}, k, 
          p_{1}, p_{2}) - 
        4 (a_{2}.p_{4}) (k.p_{1}) (k.p_{2}) (k.p_{4}) \epsilon(a_{1}, k, 
          p_{1}, p_{2}) + 
        4 (a_{2}.p_{3}) (k.p_{1})^2 (k.p_{2}) \epsilon(a_{1}, k, p_{1}, 
          p_{3}) \\&- &\nonumber 
        4 (a_{2}.p_{4}) (k.p_{1})^2 (k.p_{2}) \epsilon(a_{1}, k, p_{1}, 
          p_{3}) + 
        4 (a_{2}.p_{3}) (k.p_{1}) (k.p_{2})^2 \epsilon(a_{1}, k, p_{1}, 
          p_{3}) - 
        4 (a_{2}.p_{4}) (k.p_{1}) (k.p_{2})^2 \epsilon(a_{1}, k, p_{1}, 
          p_{3}) - 
        4 (a_{2}.p_{3}) (k.p_{1})^2 (k.p_{2})\\&\times &\nonumber \epsilon(a_{1}, k, p_{1}, 
          p_{4}) + 
        4 (a_{2}.p_{4}) (k.p_{1})^2 (k.p_{2}) \epsilon(a_{1}, k, p_{1}, 
          p_{4}) - 
        4 (a_{2}.p_{3}) (k.p_{1}) (k.p_{2})^2 \epsilon(a_{1}, k, p_{1}, 
          p_{4}) + 
        4 (a_{2}.p_{4}) (k.p_{1}) (k.p_{2})^2 \epsilon(a_{1}, k, p_{1}, 
          p_{4}) - 
        4 (a_{2}.p_{3})\\&\times &\nonumber (k.p_{1})^2 (k.p_{2}) \epsilon(a_{1}, k, p_{2}, 
          p_{3}) + 
        4 (a_{2}.p_{4}) (k.p_{1})^2 (k.p_{2}) \epsilon(a_{1}, k, p_{2}, 
          p_{3}) - 
        4 (a_{2}.p_{3}) (k.p_{1}) (k.p_{2})^2 \epsilon(a_{1}, k, p_{2}, 
          p_{3}) + 
        4 (a_{2}.p_{4}) (k.p_{1}) (k.p_{2})^2 \epsilon(a_{1}, k, p_{2}, 
          p_{3}) \\&+ &\nonumber 
        4 (a_{2}.p_{3}) (k.p_{1})^2 (k.p_{2}) \epsilon(a_{1}, k, p_{2}, 
          p_{4}) - 
        4 (a_{2}.p_{4}) (k.p_{1})^2 (k.p_{2}) \epsilon(a_{1}, k, p_{2}, 
          p_{4}) + 
        4 (a_{2}.p_{3}) (k.p_{1}) (k.p_{2})^2 \epsilon(a_{1}, k, p_{2}, 
          p_{4}) - 
        4 (a_{2}.p_{4}) (k.p_{1}) (k.p_{2})^2\\&\times &\nonumber \epsilon(a_{1}, k, p_{2}, 
          p_{4}) + 
        4 (a_{1}.p_{3}) (k.p_{1}) (k.p_{2}) (k.p_{3}) \epsilon(a_{2}, k, 
          p_{1}, p_{2}) - 
        4 (a_{1}.p_{4}) (k.p_{1}) (k.p_{2}) (k.p_{3}) \epsilon(a_{2}, k, 
          p_{1}, p_{2}) - 
        4 (a_{1}.p_{3}) (k.p_{1}) (k.p_{2}) (k.p_{4}) \epsilon(a_{2}, k, 
          p_{1}, p_{2}) \\&+ &\nonumber 
        4 (a_{1}.p_{4}) (k.p_{1}) (k.p_{2}) (k.p_{4}) \epsilon(a_{2}, k, 
          p_{1}, p_{2}) - 
        4 (a_{1}.p_{3}) (k.p_{1})^2 (k.p_{2}) \epsilon(a_{2}, k, p_{1}, 
          p_{3}) + 
        4 (a_{1}.p_{4}) (k.p_{1})^2 (k.p_{2}) \epsilon(a_{2}, k, p_{1}, 
          p_{3}) - 
        4 (a_{1}.p_{3}) (k.p_{1}) (k.p_{2})^2\\&\times &\nonumber \epsilon(a_{2}, k, p_{1}, 
          p_{3}) + 
        4 (a_{1}.p_{4}) (k.p_{1}) (k.p_{2})^2 \epsilon(a_{2}, k, p_{1}, 
          p_{3}) + 
        (a_{1}.p_{2}) (k.p_{1}) (k.p_{2}) (k.p_{3}) \epsilon(a_{2}, k, p_{1}, 
          p_{3}) - 
        (a_{1}.p_{2}) (k.p_{2})^2 (k.p_{3}) \epsilon(a_{2}, k, p_{1}, 
          p_{3}) - 
        (a_{1}.p_{2})\\&\times &\nonumber (k.p_{1}) (k.p_{2}) (k.p_{4}) \epsilon(a_{2}, k, p_{1}, 
          p_{3}) + 
        (a_{1}.p_{2}) (k.p_{2})^2 (k.p_{4}) \epsilon(a_{2}, k, p_{1}, 
          p_{3}) + 
        4 (a_{1}.p_{3}) (k.p_{1})^2 (k.p_{2}) \epsilon(a_{2}, k, p_{1}, 
          p_{4}) - 
        4 (a_{1}.p_{4}) (k.p_{1})^2 (k.p_{2})\\&\times &\nonumber \epsilon(a_{2}, k, p_{1}, 
          p_{4}) + 
        4 (a_{1}.p_{3}) (k.p_{1}) (k.p_{2})^2 \epsilon(a_{2}, k, p_{1}, 
          p_{4}) - 
        4 (a_{1}.p_{4}) (k.p_{1}) (k.p_{2})^2 \epsilon(a_{2}, k, p_{1}, 
          p_{4}) - 
        (a_{1}.p_{2}) (k.p_{1}) (k.p_{2}) (k.p_{3}) \epsilon(a_{2}, k, p_{1}, 
          p_{4}) + 
        (a_{1}.p_{2})\\&\times &\nonumber (k.p_{2})^2 (k.p_{3}) \epsilon(a_{2}, k, p_{1}, 
          p_{4}) + 
        (a_{1}.p_{2}) (k.p_{1}) (k.p_{2}) (k.p_{4}) \epsilon(a_{2}, k, p_{1}, 
          p_{4}) - 
        (a_{1}.p_{2}) (k.p_{2})^2 (k.p_{4}) \epsilon(a_{2}, k, p_{1}, 
          p_{4}) + 
        4 (a_{1}.p_{3}) (k.p_{1})^2 (k.p_{2}) \epsilon(a_{2}, k, p_{2}, 
          p_{3}) \\&- &\nonumber 
        4 (a_{1}.p_{4}) (k.p_{1})^2 (k.p_{2}) \epsilon(a_{2}, k, p_{2}, 
          p_{3}) + 
        4 (a_{1}.p_{3}) (k.p_{1}) (k.p_{2})^2 \epsilon(a_{2}, k, p_{2}, 
          p_{3}) - 
        4 (a_{1}.p_{4}) (k.p_{1}) (k.p_{2})^2 \epsilon(a_{2}, k, p_{2}, 
          p_{3}) + 
        (a_{1}.p_{1}) (k.p_{1})^2 (k.p_{3})\\&\times &\nonumber \epsilon(a_{2}, k, p_{2}, 
          p_{3}) - 
        (a_{1}.p_{1}) (k.p_{1}) (k.p_{2}) (k.p_{3}) \epsilon(a_{2}, k, p_{2}, 
          p_{3}) - 
        (a_{1}.p_{1}) (k.p_{1})^2 (k.p_{4}) \epsilon(a_{2}, k, p_{2}, 
          p_{3}) + 
        (a_{1}.p_{1}) (k.p_{1}) (k.p_{2}) (k.p_{4}) \epsilon(a_{2}, k, p_{2}, 
          p_{3}) + 
        (k.p_{1})\\&\times &\nonumber (-4 (a_{1}.p_{3}) (k.p_{2}) ((k.p_{1}) + (k.p_{2})) + 4 (a_{1}.p_{4}) (k.p_{2}) ((k.p_{1}) + (k.p_{2})) - 
           (a_{1}.p_{1}) ((k.p_{1}) - (k.p_{2})) ((k.p_{3}) - (k.p_{4}))) \epsilon(a_{2}, 
          k, p_{2}, p_{4})))\Big]
\end{eqnarray}
\normalsize
\tiny
\begin{eqnarray}
D&=&\nonumber \dfrac{2\,e}{((k.p_{1})^2 (k.p_{2})^2)}\Big[
   ((g_{a}^{e^{2}} + 
       g_{v}^{e^{2}}) (k.p_{1}) (k.p_{2}) (a^2 e^2 (((a_{1}.p_{3}) - (a_{1}.p_{4})) ((k.p_{1}) - (k.p_{2})) - ((a_{1}.p_{1}) - 
             (a_{1}.p_{2})) ((k.p_{3}) - (k.p_{4}))) ((k.p_{3}) - (k.p_{4})) \\&+ &\nonumber 
       2 (((a_{1}.p_{3}) - (a_{1}.p_{4})) (k.p_{1}) (k.p_{2}) ((p_{1}.p_{3}) - (p_{1}.p_{4}) - (p_{2}.p_{3}) + (p_{2}.p_{4})) + 
          (a_{1}.p_{1}) (k.p_{2}) (((k.p_{3}) - (k.p_{4})) ((p_{2}.p_{3}) - (p_{2}.p_{4})) - ((k.p_{1}) + (k.p_{2}))\\&\times &\nonumber (m_{A^{0}}^{2} - 
                (p_{3}.p_{4}))) + 
          (a_{1}.p_{2}) (k.p_{1}) (-((k.p_{3}) - (k.p_{4})) ((p_{1}.p_{3}) - (p_{1}.p_{4})) + ((k.p_{1}) + (k.p_{2})) (m_{A^{0}}^{2} - 
                (p_{3}.p_{4}))))) - 
    a^2 e^2 g_{a}^{e} g_{v}^{e} ((k.p_{1}) - (k.p_{2}))\\&\times &\nonumber ((k.p_{3}) - (k.p_{4}))^2 \epsilon(a_{2}, 
      k, p_{1}, p_{2}) - 
    g_{a}^{e} g_{v}^{e} (k.p_{2}) ((k.p_{1}) + (k.p_{2})) (a^2 e^2 ((k.p_{3}) - (k.p_{4})) + 
       2 (k.p_{1}) ((p_{2}.p_{3}) - (p_{2}.p_{4}))) \epsilon(a_{2}, k, 
      p_{1}, p_{3}) + 
    g_{a}^{e} g_{v}^{e} ((k.p_{2})\\&\times &\nonumber ((k.p_{1}) + (k.p_{2})) (a^2 e^2 ((k.p_{3}) - (k.p_{4})) + 
          2 (k.p_{1}) ((p_{2}.p_{3}) - (p_{2}.p_{4}))) \epsilon(a_{2}, k, 
         p_{1}, p_{4}) + 
       (k.p_{1}) (-((k.p_{1}) + (k.p_{2})) (a^2 e^2 ((k.p_{3}) - (k.p_{4})) + 
             2 (k.p_{2}) \\&\times &\nonumber((p_{1}.p_{3}) - (p_{1}.p_{4}))) \epsilon(a_{2}, k, 
            p_{2}, 
            p_{3}) + ((k.p_{1}) + (k.p_{2})) (a^2 e^2 ((k.p_{3}) - (k.p_{4})) + 
             2 (k.p_{2}) ((p_{1}.p_{3}) - (p_{1}.p_{4}))) \epsilon(a_{2}, k, 
            p_{2}, p_{4}) + 2 ((k.p_{1}) - 
             (k.p_{2}))\\&\times &\nonumber (k.p_{2}) (-((k.p_{3}) - (k.p_{4})) (\epsilon(a_{2}, p_{1}, 
                 p_{2}, p_{3}) - 
                \epsilon(a_{2}, p_{1}, p_{2}, 
                 p_{4})) + ((a_{2}.p_{3}) - 
                (a_{2}.p_{4})) (\epsilon(k, p_{1}, p_{2}, 
                 p_{3}) - 
                \epsilon(k, p_{1}, p_{2}, 
                 p_{4}))))))\Big]
\end{eqnarray}
\normalsize
\tiny
\begin{eqnarray}
E&=&\nonumber\dfrac{2\,e}{((k.p_{1})^2 (k.p_{2})^2)}\Big[
   ((g_{a}^{e^{2}} + 
       g_{v}^{e^{2}}) (k.p_{1}) (k.p_{2}) (a^2 e^2 (((a_{1}.p_{3}) - (a_{1}.p_{4})) ((k.p_{1}) - (k.p_{2})) - ((a_{1}.p_{1}) - 
             (a_{1}.p_{2})) ((k.p_{3}) - (k.p_{4}))) ((k.p_{3}) - (k.p_{4})) \\&+ &\nonumber 
       2 (((a_{1}.p_{3}) - (a_{1}.p_{4})) (k.p_{1}) (k.p_{2}) ((p_{1}.p_{3}) - (p_{1}.p_{4}) - (p_{2}.p_{3}) + (p_{2}.p_{4})) + 
          (a_{1}.p_{1}) (k.p_{2}) (((k.p_{3}) - (k.p_{4})) ((p_{2}.p_{3}) - (p_{2}.p_{4})) - ((k.p_{1}) \\&+ &\nonumber(k.p_{2})) (m_{A^{0}}^{2} - 
                (p_{3}.p_{4}))) + 
          (a_{1}.p_{2}) (k.p_{1}) (-((k.p_{3}) - (k.p_{4})) ((p_{1}.p_{3}) - (p_{1}.p_{4})) + ((k.p_{1}) + (k.p_{2})) (m_{A^{0}}^{2} - 
                (p_{3}.p_{4}))))) + 
    a^2 e^2 g_{a}^{e} g_{v}^{e} ((k.p_{1}) \\&- &\nonumber (k.p_{2})) ((k.p_{3}) - (k.p_{4}))^2 \epsilon(a_{2}, 
      k, p_{1}, p_{2}) + 
    g_{a}^{e} g_{v}^{e} (k.p_{2}) ((k.p_{1}) + (k.p_{2})) (a^2 e^2 ((k.p_{3}) - (k.p_{4})) + 
       2 (k.p_{1}) ((p_{2}.p_{3}) - (p_{2}.p_{4}))) \epsilon(a_{2}, k, 
      p_{1}, p_{3}) \\&+ &\nonumber
    g_{a}^{e} g_{v}^{e} (-(k.p_{2}) ((k.p_{1}) + (k.p_{2})) (a^2 e^2 ((k.p_{3}) - (k.p_{4})) + 
          2 (k.p_{1}) ((p_{2}.p_{3}) - (p_{2}.p_{4}))) \epsilon(a_{2}, k, 
         p_{1}, p_{4}) + 
       (k.p_{1}) (((k.p_{1}) + (k.p_{2})) (a^2 e^2 ((k.p_{3}) \\&- &\nonumber(k.p_{4})) + 
             2 (k.p_{2}) ((p_{1}.p_{3}) - (p_{1}.p_{4}))) \epsilon(a_{2}, k, 
            p_{2}, 
            p_{3}) - ((k.p_{1}) + (k.p_{2})) (a^2 e^2 ((k.p_{3}) - (k.p_{4})) + 
             2 (k.p_{2}) ((p_{1}.p_{3}) - (p_{1}.p_{4}))) \epsilon(a_{2}, k, 
            p_{2}, p_{4}) \\&+ &\nonumber 2 ((k.p_{1}) - 
             (k.p_{2})) (k.p_{2}) (((k.p_{3}) - 
                (k.p_{4})) (\epsilon(a_{2}, p_{1}, p_{2}, 
                 p_{3}) - 
                \epsilon(a_{2}, p_{1}, p_{2}, 
                 p_{4})) - ((a_{2}.p_{3}) - 
                (a_{2}.p_{4})) (\epsilon(k, p_{1}, p_{2}, 
                 p_{3}) - 
                \epsilon(k, p_{1}, p_{2}, 
                 p_{4}))))))\Big]
\end{eqnarray}
\normalsize
\tiny
\begin{eqnarray}
F&=&\nonumber\dfrac{4 e^2}{((k.p_{1}) (k.p_{2}))}
    \Big[(g_{a}^{e^{2}} + g_{v}^{e^{2}}) ((a_{1}.p_{3})^2 (k.p_{1}) (k.p_{2}) + 
     (a_{1}.p_{4})^2 (k.p_{1}) (k.p_{2}) - ((a_{2}.p_{3}) - (a_{2}.p_{4}))^2 (k.p_{1}) (k.p_{2}) + 
     (a_{1}.p_{3}) (-2 (a_{1}.p_{4}) (k.p_{1})\\&\times &\nonumber (k.p_{2}) - ((a_{1}.p_{2}) (k.p_{1}) + (a_{1}.p_{1}) (k.p_{2})) ((k.p_{3}) - (k.p_{4}))) + 
     (a_{1}.p_{4}) ((a_{1}.p_{2}) (k.p_{1}) + (a_{1}.p_{1}) (k.p_{2})) ((k.p_{3}) - (k.p_{4})) + (a_{1}.p_{1})\\&\times &\nonumber (a_{1}.p_{2}) ((k.p_{3}) - (k.p_{4}))^2)\Big]
\end{eqnarray}

\end{document}